\documentclass[prd, aps, twocolumn, showpacs,superscriptaddress,nofootinbib]{revtex4-1}

\usepackage{amsbsy,amsmath,amssymb,amsthm,wasysym,amsfonts}
\usepackage{graphicx}
\usepackage{psfrag}
\usepackage{epstopdf}
\usepackage{color}
\usepackage{mathrsfs}
\usepackage{comment}
\usepackage{booktabs}
\usepackage{multirow}
\usepackage[normalem]{ulem}
\DeclareSymbolFontAlphabet{\mathrsfs}{rsfs}
\DeclareMathAlphabet\mathbfcal{OMS}{cmsy}{b}{n}

\newcommand{\be}{\begin{equation}}  
\newcommand{\ee}{\end{equation}}
\newcommand{\bea}{\begin{eqnarray}}           
\newcommand{\eea}{\end{eqnarray}} 
\newcommand{\beqn}{\begin{eqnarray*}}
\newcommand{\eeqn}{\end{eqnarray*}}
\newcommand{\ba}{\begin{align}}
\newcommand{\ea}{\end{align}}

\def\f{{\hat{f}}}
\def\p4{{\psi_4}} 

\usepackage{float}
\definecolor{cyan}{rgb}{0,0.9,0.9}
\definecolor{orange}{rgb}{0.9,0.5,0}
\definecolor{magenta}{rgb}{1,0,1}
\definecolor{purple}{rgb}{0.8,0.4,0.8}

\definecolor{dodgerblue}{rgb}{0.12, 0.56, 1.0}
\definecolor{alizarincrimson}{rgb}{0.82, 0.1, 0.26}

\begin{document}

\title{Quasi-5.5PN TaylorF2 approximant for compact binaries: \\
  point-mass phasing and impact on the tidal polarizability inference}

\author{Francesco \surname{Messina}}
\affiliation{Dipartimento di Fisica, Universit\`a degli studi di Milano Bicocca, Piazza della Scienza 3, 20126 Milano, Italy}
\affiliation{INFN, Sezione di Milano Bicocca, Piazza della Scienza 3, 20126 Milano, Italy}
\author{Reetika \surname{Dudi}}
\affiliation{Theoretisch-Physikalisches Institut, Friedrich-Schiller-Universit{\"a}t Jena, 07743, Jena, Germany}  
\author{Alessandro \surname{Nagar}}
\affiliation{Centro Fermi - Museo Storico della Fisica e Centro Studi e Ricerche Enrico Fermi, Rome, Italy}
\affiliation{INFN Sezione di Torino, Via P. Giuria 1, 10125 Torino, Italy}
\affiliation{Institut des Hautes Etudes Scientifiques, 91440 Bures-sur-Yvette, France}
\author{Sebastiano \surname{Bernuzzi}}
\affiliation{Theoretisch-Physikalisches Institut, Friedrich-Schiller-Universit{\"a}t Jena, 07743, Jena, Germany}

\begin{abstract}
We derive a point-mass (nonspinning) frequency-domain {\tt TaylorF2} phasing
approximant at {\it quasi}-5.5~post-Newtonian (PN) accuracy for the
gravitational wave from coalescing compact binaries. The new
approximant is obtained by Taylor-expanding the effective-one-body
(EOB) resummed energy and and angular momentum flux along circular
orbits with all the known test-particle information up to 5.5PN. The
-yet uncalculated- terms at 4PN order and beyond entering both the
energy flux and the energy are taken into account as free parameters
and then set to zero.
We compare the quasi-5.5PN and 3.5PN approximants against full EOB
waveforms using gauge-invariant phasing diagnostics
$Q_\omega=\hat\omega^2/\dot{\hat\omega}$, where $\hat\omega$ is the
dimensionless gravitational-wave frequency.
The quasi-5.5PN phasing is found to be systematically closer to the EOB one 
than the 3.5PN one. Notably, the quasi-5.5PN (3.5PN) approximant accumulates
a EOB$-$PN dephasing of $\Delta\Psi^{\rm EOBPN}\sim10^{-3}$~rad
($0.13$~rad) up to frequency $\hat\omega \simeq 0.06$, 6 orbits to merger, 
($\hat\omega \simeq 0.086$, 2 orbits to merger) for a fiducial binary
neutron star system. 
We explore the performance of the quasi-5.5PN approximant on
the measurement of the tidal polarizability parameter $\tilde\Lambda$
using injections of EOB waveforms hybridized with numerical relativity
merger waveforms.
We prove that the quasi-5.5PN point-mass approximant augmented with
6PN-accurate tidal terms allows one to reduce (and in many cases even
eliminate) the biases in the measurement of $\tilde\Lambda$ that are
instead found when the standard 3.5PN point-mass baseline is used.
Methodologically, we demonstrate that the combined use of $Q_\omega$ analysis
and of the Bayesian parameter estimation offers a new tool to investigate the
impact of systematics on gravitational-wave inference.
\end{abstract}

\maketitle

\section{Introduction}
The data-analysis of GW170817~\cite{TheLIGOScientific:2017qsa} relied on gravitational
waveform models that incorporate tidal effects. The latter allow one
to extract information about the neutron star equation of state (EOS)
via the inference of the mass-weighted averaged tidal polarizability
parameter $\tilde\Lambda$
\cite{Damour:2012yf,DelPozzo:2013ala,Abbott:2018exr,Abbott:2018wiz,TheLIGOScientific:2018mvr}. 
The understanding of the systematic uncertainties on the measurement of 
$\tilde{\Lambda}$ due to the waveform model/approximants have been the subject of intensive investigation in recent years.
For example, building on the work of Favata~\cite{Favata:2013rwa}, 
Wade et al.~\cite{Wade:2014vqa} investigated the performance of
different PN inspiral approximant within a Bayesian analysis framework for the advanced 
detectors and found that the choice of approximant significantly  
biases the recovery of tidal parameters. Later, a similar Bayesian analysis in the case of LIGO 
and advanced LIGO detectors was carried out by Dudi et al.~\cite{Dudi:2018jzn} who 
concluded that the {\tt TaylorF2} 3.5PN waveform model can be used to place an upper 
bound on $\tilde{\Lambda}$. The same conclusion was drawn also by the 
study of the LIGO-Virgo collaboration~\cite{Abbott:2018wiz}.

Beside interesting per se' because done in the precise
setup that is relevant for data analysis, these studies collectively stress 
the paramount need of having an analytically reliable description of the 
phasing up to merger. The tidal extension of the effective-one-body 
(EOB)~\cite{Buonanno:1998gg,Buonanno:2000ef} description for coalescing compact binaries was introduced
in~\cite{Damour:2009wj} and developed during the last ten
years~\cite{Bernuzzi:2014owa,Hinderer:2016eia,Steinhoff:2016rfi,Nagar:2018zoe,Akcay:2018yyh,Nagar:2018plt} 
with the goal of providing robust binary neutron stars (BNS) waveforms to be
used in gravitational-wave inference.
While analytically more accurate, EOB waveform generation is usually
slower than PN. Different routes have been explored to speedup EOB approximants.
One possibility is to construct surrogate waveform models~\cite{Lackey:2016krb,Lackey:2018zvw}.
Another possibility is to conjugate the efficiency of a PN approximant
with the physical completeness of the EOB model; the {\tt NRTidal}
family of approximants partly answers to this question~\cite{Dietrich:2017aum,Dietrich:2018uni}. 
A third approach relies on the development of fast approximations for
the solution of EOB equations, for example using the high-order
post-adiabatic approach of Ref.~\cite{Nagar:2018gnk} (See also \cite{Akcay:2018yyh,Nagar:2018plt}.)
However, none of the methods described above provide us with waveform
generation algorithms faster than PN. Although it is well known that
inspiral PN approximants might be problematic, they retain the
advantage of being the most efficient for Bayesian inference. 

One important source of systematics in BNS inspiral waveforms resides in 
the description of the nontidal part (see e.g.~\cite{Samajdar:2018dcx}).
The practice that became common after the observation of GW170817 is to augment
standard point-mass model with the tidal part of the phasing.
A natural step is thus to improve the accuracy point-mass PN
approximant beyond the current available. 
In this paper we introduce a nonspinning,
point-mass, closed-form frequency-domain {\tt TaylorF2} waveform approximant at
quasi-5.5PN order (Sec.~\ref{sec:5p5PNorbQomg}).
The new approximant is obtained by PN-expanding the adiabatic
EOB dynamics along circular orbits. As such, it delivers a phasing
representation that improves the currently known 3.5PN one.
We show that, when applied in the GW data analysis context, the new phasing description allows
one to strongly reduce the biases in the recovery of the tidal
parameters that are usually present with the 3.5PN {\tt TaylorF2} point-mass 
(Sec.~\ref{sec:DA}).

In the following, the total gravitational mass of
the binary is $M=m_1+m_2$, with the two bodies labeled as $(1,2)$.
We adopt the convention that that $m_1\geq m_2$, so to define
the mass ratio $q\equiv m_1/m_2\geq 1$ and the symmetric mass ratio
$\nu\equiv m_1 m_2/M^2$, that ranges from $0$ (test-particle limit) to
$\nu=1/4$ (equal-mass case). The dimensionless spins are addressed
as $\chi_{i}\equiv S_{i}/m_{i}^2$. We also define the quantities 
$X_{i}\equiv m_i/M$ and $X_{12}\equiv X_{1}-X_{2}=\sqrt{1-4\nu}$,
which yields $X_1=\frac{1}{2}(1+\sqrt{1-4\nu})$ and $X_2=1-X_1$. 
Following Refs.~\cite{Nagar:2018zoe,Messina:2017yjg,Nagar:2018plt}, 
it is also convenient to use the following spin variables or spin-related 
quantities: $\tilde{a}_{i}\equiv S_i/(m_i M)=X_{i}\chi_{i}$;
$\tilde{a}_0\equiv \tilde{a}_1+\tilde{a}_2$ and $\tilde{a}_{12}\equiv \tilde{a}_1-\tilde{a}_2$.
If not otherwise specified, we use geometric units $c=G=1$. To convert
from geometric to physical units we recall that
$GM_\odot/c^3=4.925491\times 10^{-6}$~sec.

\section{Quasi-5.5PN-accurate orbital phasing}
\label{sec:5p5PNorbQomg}

Building upon Damour et al.~\cite{Damour:2012yf}, Ref.~\cite{Messina:2017yjg}  illustrated 
how to formally obtain a high-order PN approximant by PN-expanding the EOB energy $E_{\rm EOB}$ 
and energy flux ${\cal F}_{\rm EOB}$ along circular  orbits. Stopping
the expansion at 4.5PN, allowed one to obtain  a consistent 4.5PN
approximant with a few parameters needed to formally take into account  
the yet uncalculated $\nu$-dependent terms in the waveform amplitudes at 4PN. Here we follow
precisely that approach, but we extend it to 5.5PN accuracy. To get the waveform phase in the 
frequency domain along circular orbits, we start with the
gauge-invariant\footnote{In the sense that it is independent of
  time and phase arbitrary shifts.} description of the adiabatic
phasing given by the function
\begin{equation}
  \label{eq:defQ}
  Q_\omega\equiv E_{\rm EOB}(x)\left(\dfrac{ d{\cal F}_{\rm EOB}}{dx}\right)^{-1} \ ,
\end{equation}
where $x\equiv (M\Omega)^{2/3}$, with $\Omega$ the orbital frequency along circular orbits.
The high-order phasing approximant is obtained by Taylor-expanding the above equation
and then by solving the equation
\begin{equation}\label{eq:Psi}
\frac{d^2\Psi_{\rm 5.5PN}}{d^2\f}=\dfrac{Q_\omega(\f)}{\f^2} \ ,
\end{equation}
where $\hat{f}\equiv Mf \equiv \Omega /2\pi$.
The double integration of Eq.~\eqref{eq:Psi} delivers $\Psi_{\rm
  5.5PN}(\f)$ modulo an affine part of the  
form $p+q\hat{f}$, where $(p,q)$ are two arbitrary integration constants that are
fixed to be consistent with the usual conventions adopted in the literature for the
3.5~PN approximant~\cite{Damour:2000zb}. 

We consider here only
{\it nonspinning} binaries (the reader is referred to Appendix~\ref{sec:mass_ratio} for the discussion of the spin case). The corresponding, circularized, EOB Hamiltonian reads
$H^{\rm EOB}_0=M\sqrt{1+2\nu(\hat{H}_{\rm eff}-1)}$ where  $\hat{H}_{\rm eff}\equiv H_{\rm eff}/\mu=\sqrt{A(u)(1+u^2 j^2)}$,
where $j$ is the orbital angular momentum along circular orbits, $u\equiv M/R$ the inverse radial
separation and $A(u)$ is the EOB interaction potential kept with a 5PN term $\nu(a_6^c+a_6^{\ln}\ln u)u^6$
with $a_6^c$ an analytically unknown coefficient. The orbital angular momentum along circular orbits
$j$ is obtained solving  $\partial_u\hat{H}_{\rm eff}=0$. By PN-expanding one of Hamilton's
equations, $M\Omega=\partial_\varphi H^{\rm EOB}$, one obtains $x(u)$ as a 5PN truncated 
series in $u$, that, once inverted, allows to  obtain the (formal) 5.5~PN accurate energy flux 
as function of $x$ by PN-expanding its general EOB expression
$\mathcal{F}=\sum_{\ell=2}^{\infty}\sum_{\ell =-m}^{m} F^{\rm Newt}_{\ell m}\hat{F}_{\ell m}$,
where  $F^{\rm Newt}_{\ell m}$ is the Newtonian (leading-order) contribution and 
$\hat{F}_{\ell m}$ is the relativistic correction. Each multipolar contribution within the EOB formalism 
comes written in factorized and resummed form as
\begin{equation}
\label{eq:hatF}
\hat{F}_{\ell m} = \left(\hat{S}^{(\epsilon)}_{\rm eff}\right)^{2}|T_{\ell m}|^{2}(\rho_{\ell m})^{2\ell}.
\end{equation}
Here, $\hat{S}^{(\epsilon)}_{\rm eff}$ is the effective source, that is the effective
EOB energy $\hat{E}_{\rm eff}(x)\equiv E_{\rm eff}/\mu$ when $\epsilon=0$  ($\ell+m$=even)
or the Newton-normalized orbital angular  momentum when $\epsilon=1$ ($\ell+m$=odd).
The square modulus of the tail factor $T_{\ell m}$ resums an infinite number of 
PN hereditary logarithms~\cite{Damour:2008gu,Faye:2014fra}. We use the relativistic residual
amplitude ($\rho_{\ell m}$) information reported in Eqs (7)-(18) of Ref.~\cite{Messina:2017yjg},
where the unknown high-PN coefficients (polynomials in $\nu$) have been parametrized by
some coefficients $c_{\ell m}$. We include for consistency all the
coefficients to go up to the $\ell=7$, $m={\rm even}$ multipoles.

From Eq.~\eqref{eq:defQ} one obtains the following PN-expanded
expression
\begin{align}
 \label{eq:Qw5p5pn}
  &\hat{Q}_{\omega}^{\rm PN}=1+b_2x+b_3x^{3/2}+b_4x^2+b_5x^{5/2}\nonumber\\
                         &+b_6x^3+b_7x^{7/2}+b_8x^4+b_9x^{9/2}+b_{10}x^5+b_{11}x^{11/2}.
\end{align}
The coefficients of this expansion, that are reported in full  in Appendix~\ref{sec:coefficients}, 
have the structure $b_i\equiv b_i^0+b_i(\nu)$, where
$b_i^0$ is the $\nu$-independent (test-particle) part, fully known analytically,
while the $b_i(\nu)$ encode the $\nu$-dependence that is completely known at 3PN,
while only partially known at 4PN because the corresponding waveform calculation
is not completed yet. The $\nu$-dependence beyond 3PN is formally incorporated
by extending the analytically known $\rho_{\ell m}$ function with additional
$\nu$-dependent coefficients and then reflects in the coefficients $b_i(\nu)$.
Among these coefficients, those that depend on the parameters that we have 
introduced in the computation are
\begin{align}
b_8 & =b_8(c_{21}^{\rm 3PN},c_{22}^{\rm 4PN}) \\
b_{10} & =b_{10}(c_{21}^{\rm 3PN},c_{22}^{\rm 4PN},c_{22}^{\rm 5PN}) \\
b_{11} &=b_{11}(c_{21}^{\rm 3PN},c_{22}^{\rm 4PN}) \ .
\end{align}
In the following analysis,
we fix to zero $a_6^c$ as well as all the yet uncalculated, $\nu$-dependent, PN 
waveform coefficients entering Eq.~\eqref{eq:Qw5p5pn} above. This entitles us
to use the definition of {\it quasi-5.5PN} approximant (this PN-order choice is discussed in Appendix \ref{sec:why_5p5PN} and resumed in Fig. \ref{fig:Qwcomp_q1_ALL}). Note however that in the 
NR-informed EOB model, that we shall use to check the reliability of this 
quasi-5.5PN approximant, all the waveform coefficients are equally fixed to be
zero; on the contrary, $a_6^c$ is informed by NR simulations and, as such,
effectively takes into account, to same extent, all this missing analytical 
information. The importance of the $\nu$-dependent 
waveform coefficients is, a priori, expected to be low, as suggested 
in Table II of~\cite{Messina:2018ghh}. This is in accord with the fact
that an eventual tuning of some free parameters is better when they
tend to be small (see Appendix \ref{sec:why_5p5PN}).  

\subsection{Assessing the 5.5PN phasing accuracy}

\begin{figure}[t]
\center
  \includegraphics[width=0.90\columnwidth]{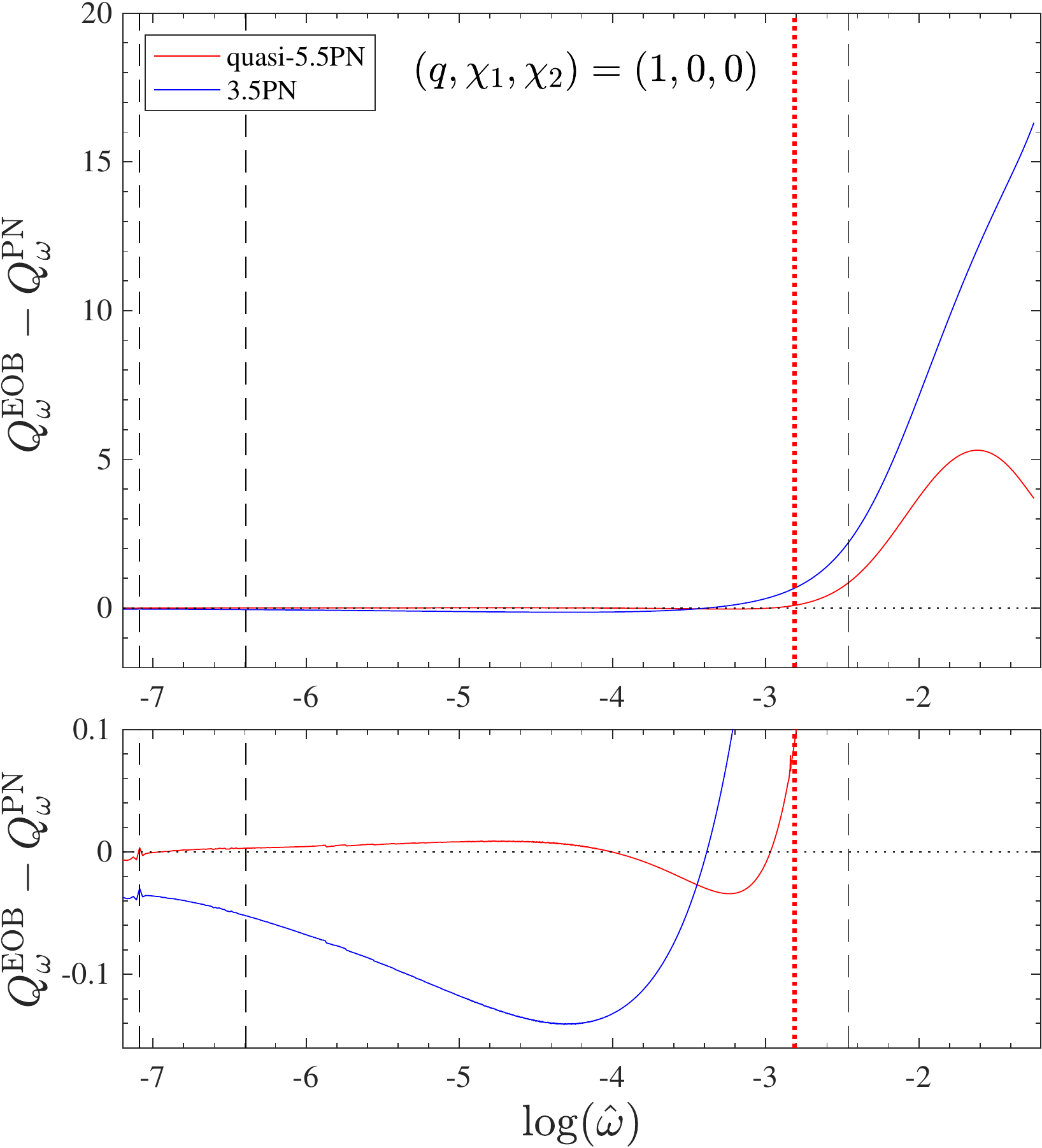} 
  \caption{ \label{fig:Qwcomp_q1} Comparison between the point-mass (nonspinning) orbital phasing for $q=1$
$Q_{\omega}^{\rm EOB}-Q_{\omega}^{\rm PN}$ difference up to (approximate) merger time. The vertical lines mark
the 10Hz, 20Hz, 718Hz or 1024Hz for a  $(1.35+1.35)M_{\odot}$ binary. The quasi-5.5PN curve is always
much closer to the EOB one than the standard 3.5PN approximant.}
\end{figure}

Let us now study the performance of {\tt TaylorF2} at 5.5PN versus the full EOB phasing. 
We do so by comparing the corresponding $Q_\omega$ functions and taking the differences,
similarly to what was done in Ref.~\cite{Dietrich:2018uni} for isolating the tidal part of the
EOB phasing and in Ref.~\cite{Nagar:2018plt} for isolating the quadratic-in-spin part. 
Since we have in mind an application to a paradigmatic BNS system, Fig.~\ref{fig:Qwcomp_q1} 
only focuses on the $q=1$ case. We show our results in terms of differences between the EOB
and the PN curves, $Q_\omega^{\rm EOB}-Q_\omega^{\rm PN}$. The top panel of the figure illustrates
the full phasing acceleration evolution, up to the peak of the EOB
orbital frequency that is 
identified with the merger. The bottom panel is a close up on the inspiral part.
The vertical lines corresponds (from left to right) to 10Hz, 20Hz, 718Hz and 1024Hz for a fiducial
equal-mass BNS system with $(1.35+1.35)M_\odot$. The 718Hz line
corresponds to $\hat{\omega}=0.06$, that roughly corresponds to the NR
contact frequency~\cite{Bernuzzi:2012ci}.
The figure highlights that the quasi-5.5PN approximant delivers a rather good  representation 
of the point-mass EOB phasing precisely up to $\hat\omega=0.06$.
Table~\ref{tab:Dphi} reports the phase difference 
\begin{equation}
\Delta
\phi_{(\hat{\omega}_0,\hat{\omega}_1)}=\int_{\hat{\omega}_0}^{\hat{\omega}_1}\Delta
Q_\omega d\log\hat{\omega} \ ,
\end{equation}
accumulated between
the frequencies $[\hat\omega_0,\hat\omega_1]$ (or equivalently
$[f_0,f_1]$ in physical units)
marked by vertical lines in the plots. The numbers in the table illustrate quantitatively how the 
5.5PN phasing approximant delivers  a phasing description that is, by itself, more 
EOB compatible than the standarly used 3.5PN one. 
Note that this is achieved even if the EOB incorporates the effective, NR-informed, 
$a_c^6(\nu)$ parameter, that is not included in the {\tt TaylorF2} approximant.

\begin{table}
  \caption{\label{tab:Dphi} EOB/PN phase difference accumulated between $[f_0,f_1]$. It is 
  obtained by integrating the $\Delta Q_\omega^{\rm EOBPN}$'s in Fig.~\ref{fig:Qwcomp_q1} 
  between the corresponding values of $\log(\hat{\omega})$. The limits of integration 
  are denoted in Hz as we want to ideally refer to the fiducial $(1.35+1.35)M_\odot$ binary system.}
  \begin{ruledtabular}
    \begin{tabular}{ccccccc}
      $\hat\omega_0\times 10^4$  & $\hat\omega_1$ & $f_0$[Hz]  & $f_1$[Hz]  & $\Delta\phi^{\rm EOBPN}_{\rm 3.5PN}$ & $\Delta\phi^{\rm EOBPN}_{\rm 5.5PN}$ \\
      \hline
      \hline
      8.35 & 0.086 & 10 & 1024 & 0.2718 & 0.1364 \\
      8.35 & 0.060 & 10 & 718   & \!\!\!\!$-0.1916$ & $1.45\times 10^{-3}$ \\
      \hline
      16.7 & 0.086 & 20 & 1024 & 0.3009 & 0.1354 \\
      16.7 & 0.060 & 20 & 718   & \!\!\!\!$-0.1625$ & $4.54\times 10^{-4}$\\
      \hline
      20.0 & 0.086 & 24 & 1024 & 0.3110  & 0.1348 \\            
    \end{tabular}
  \end{ruledtabular}
  \end{table}

\section{Application to $\tilde\Lambda$ inference}
\label{sec:DA}

\begin{figure}[t]
\begin{center}
\includegraphics[width=0.9\columnwidth]{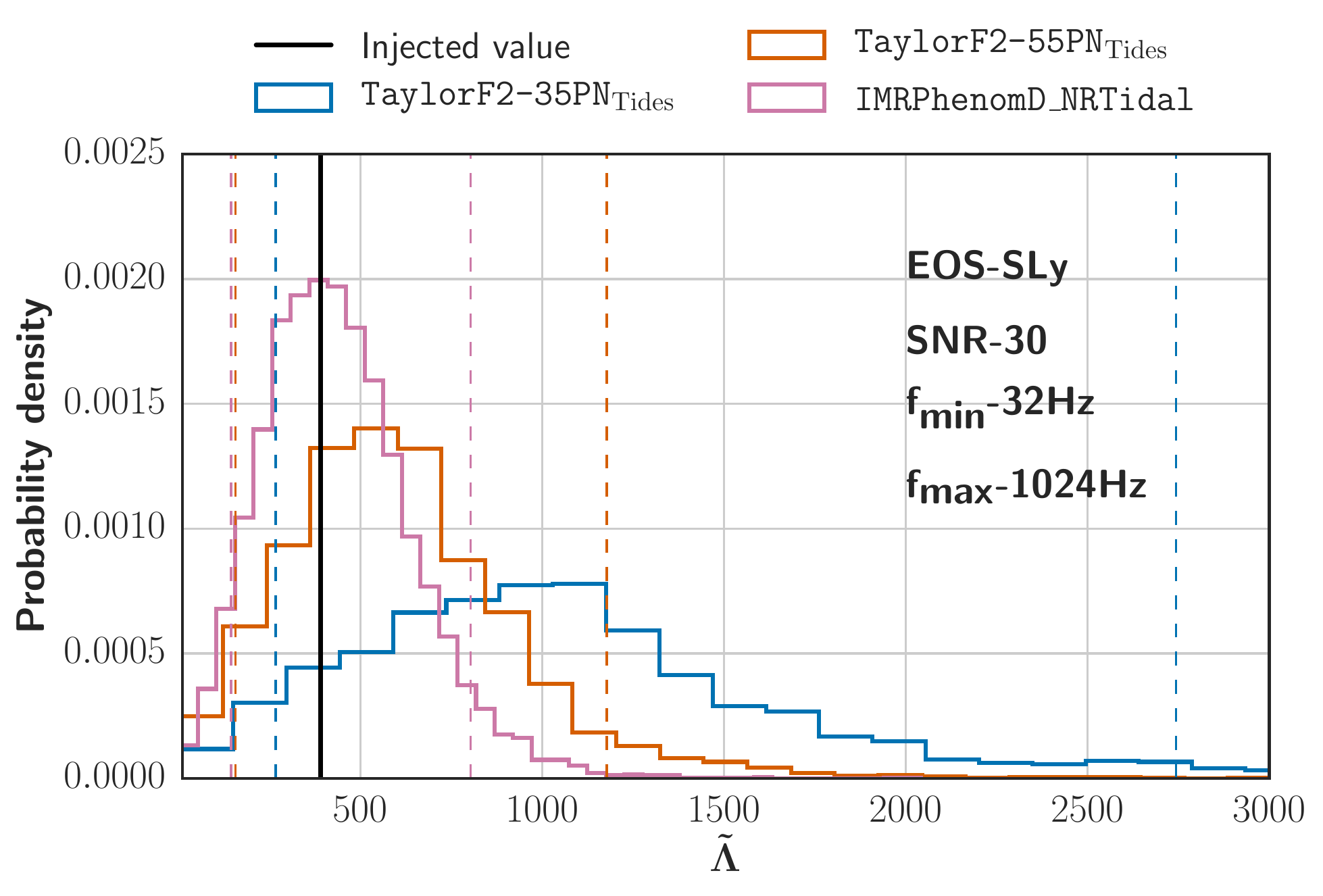}
\includegraphics[width=0.9\columnwidth]{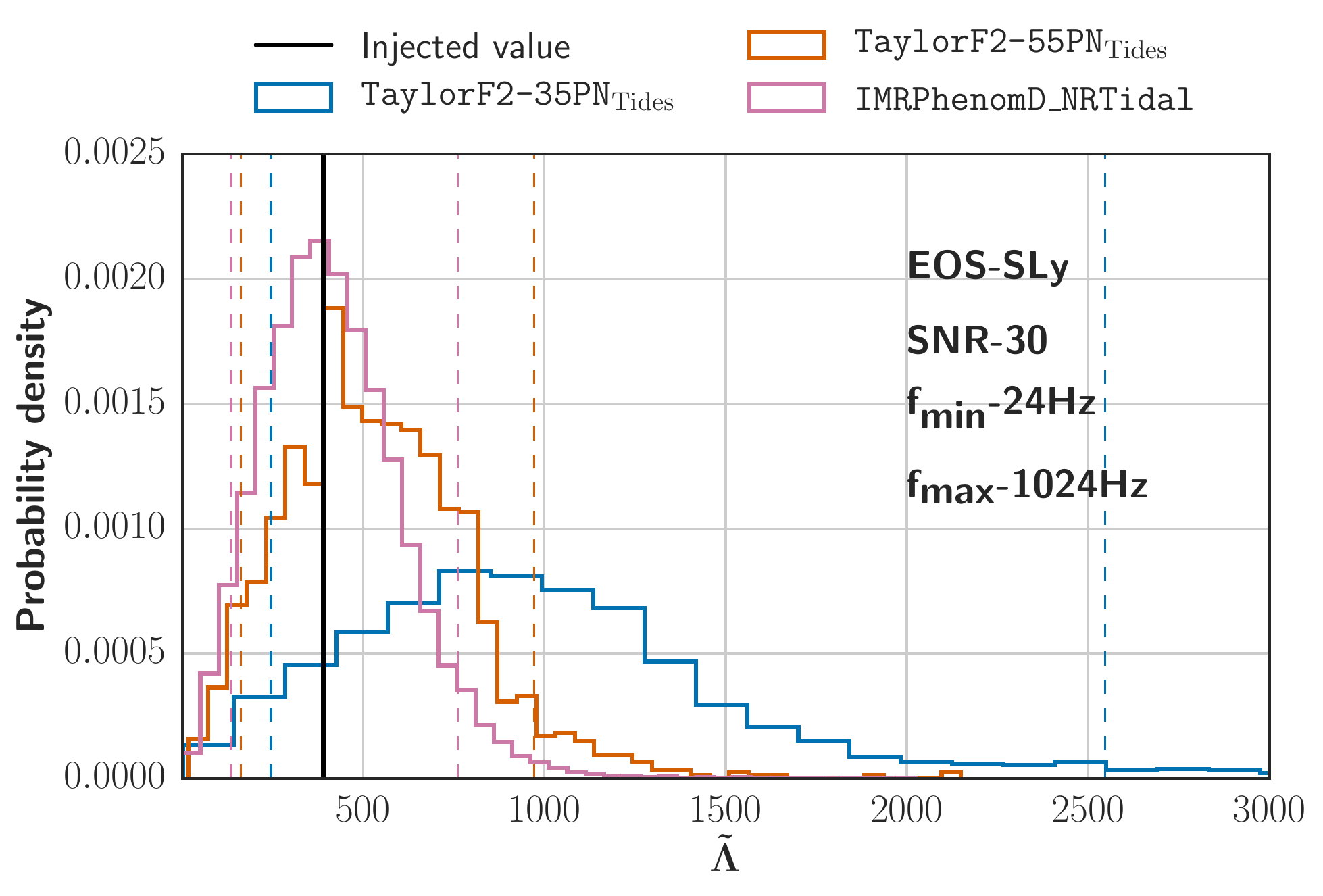}
\includegraphics[width=0.9\columnwidth]{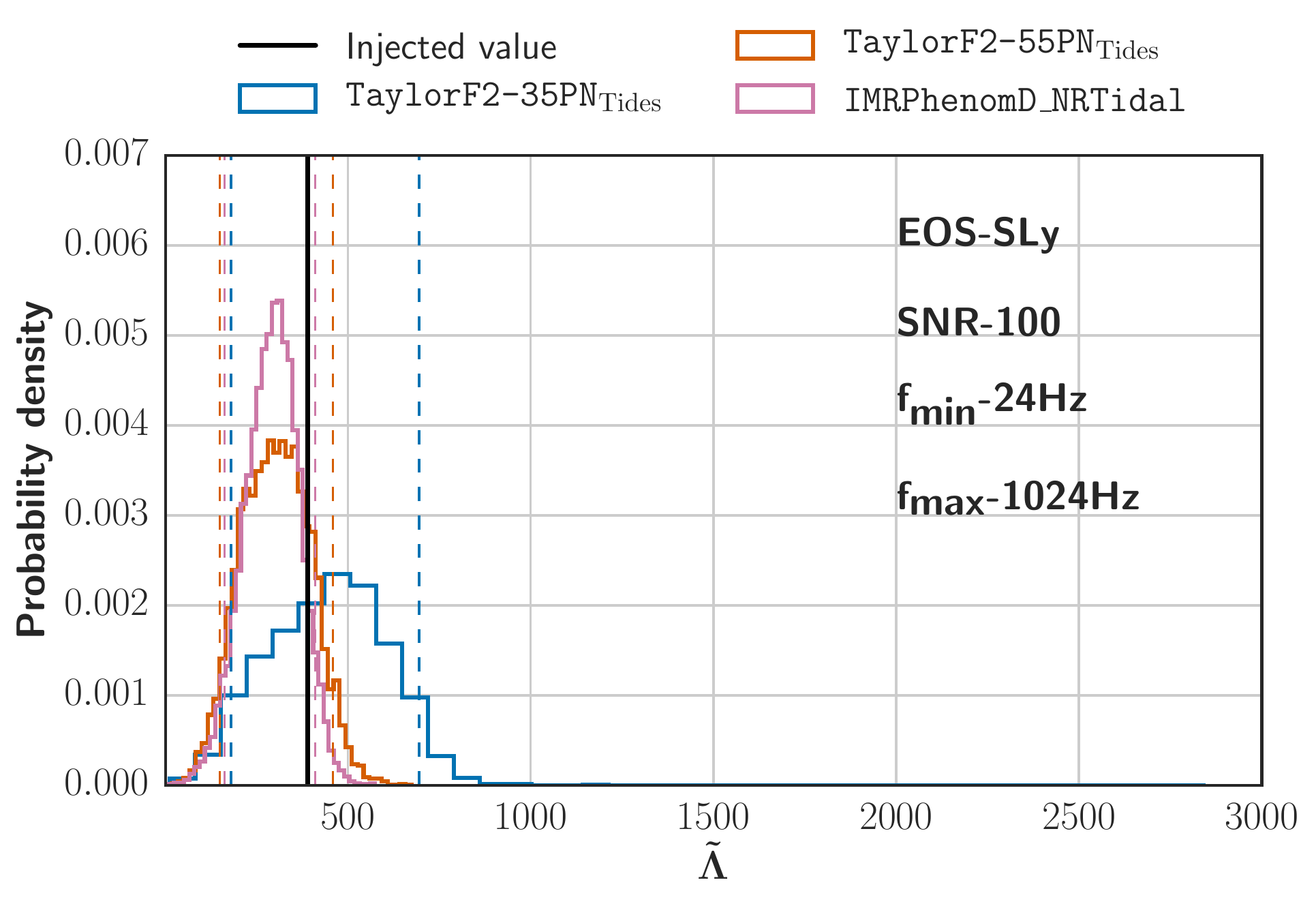}
\end{center}
\caption{ \label{fig:sly}$1.35M_\odot+1.35M_\odot$ binary with Sly EOS.
  Inference of $\tilde{\Lambda}$ with different waveform model on different
  frequency intervals $[f_{\rm min},f_{\rm max}]$ with different SNR.
  The vertical line corresponds to the injected value $\tilde{\Lambda}^{\rm SLy}=392.231$.
  Irrespectively of the value of SNR, the 3.5PN baseline introduces a
  strong bias (and spread) in the measure of $\tilde{\Lambda}$.
  By contrast this is practically reabsorbed when using the quasi-5.5PN
  point-mass baseline. The dashed vertical lines corresponds to $90\%$ confidence level.}
\end{figure}

\begin{figure}[t]
\begin{center}
\includegraphics[width=0.9\columnwidth]{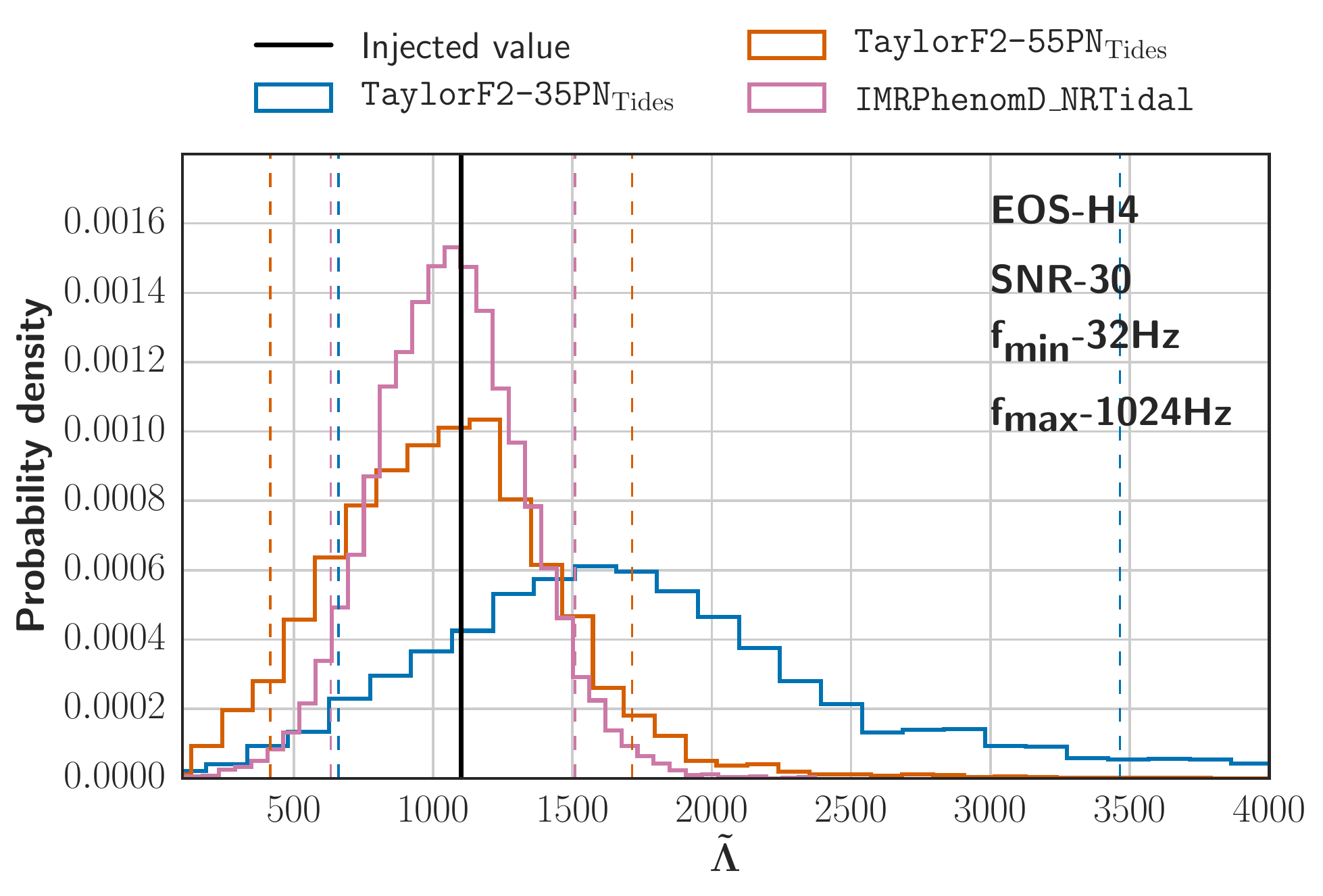}
\includegraphics[width=0.9\columnwidth]{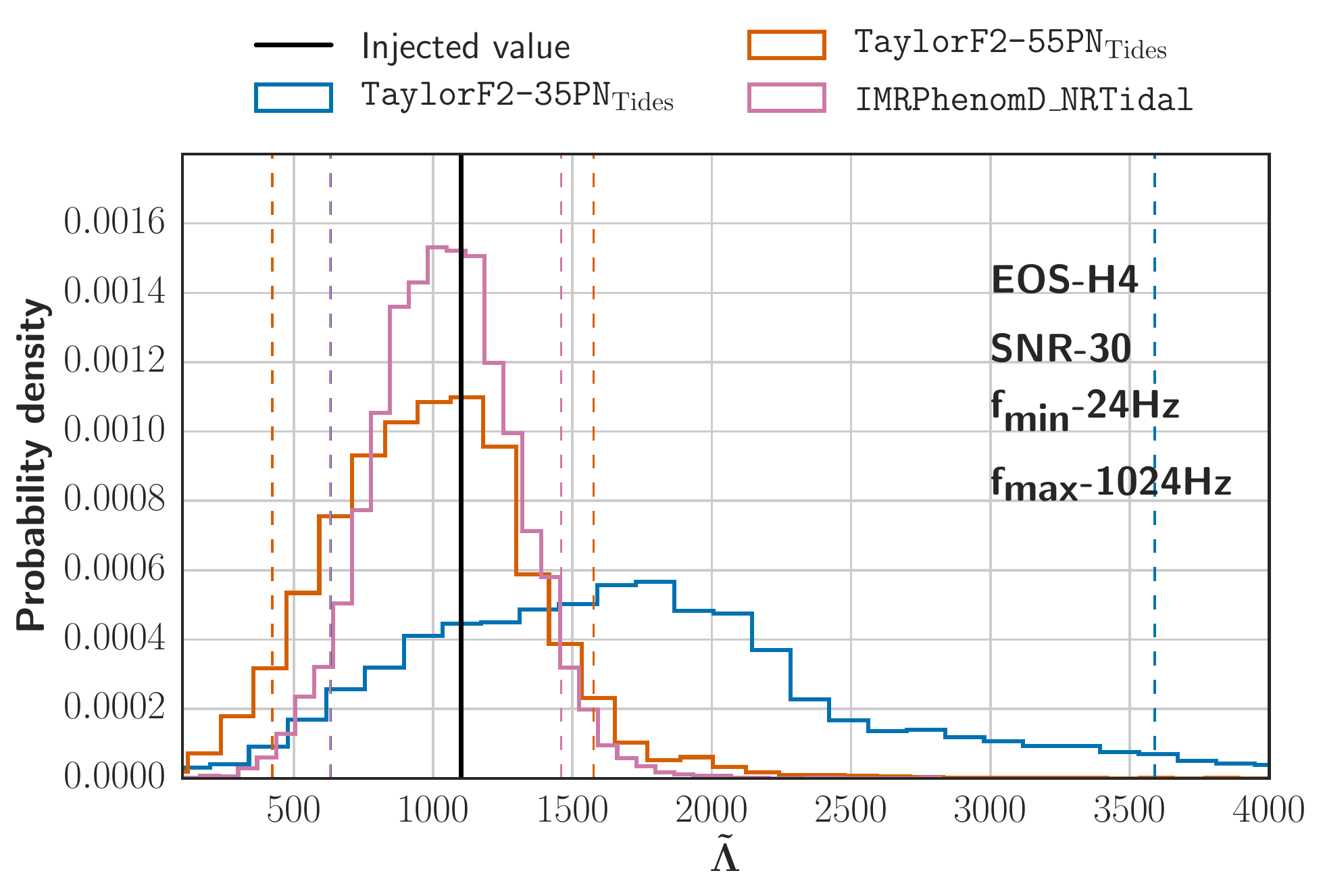}
\includegraphics[width=0.9\columnwidth]{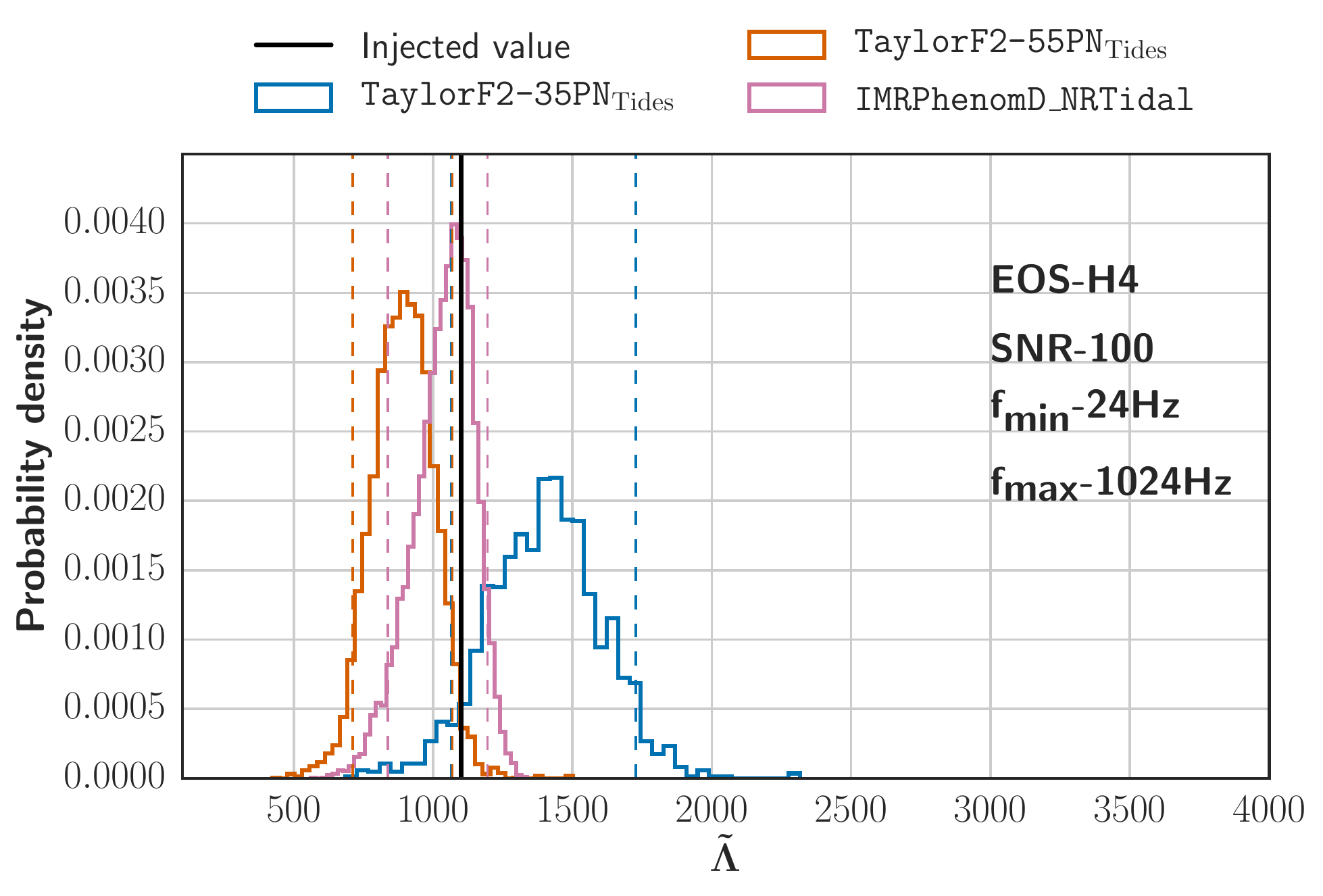}
\end{center}
\caption{ \label{fig:H4}$1.35M_\odot+1.35M_\odot$ binary with Sly EOS.
  Inference of $\tilde{\Lambda}$ with different waveform model on different
  frequency intervals $[f_{\rm min},f_{\rm max}]$ with different SNR.
  The vertical line corresponds to the injected value $\tilde{\Lambda}^{\rm SLy}=1020.5$.
  Irrespectively of the value of SNR, the 3.5PN baseline introduces a
  strong bias (and spread) in the measure of $\tilde{\Lambda}$.
  By contrast this is practically reabsorbed when using the quasi-5.5PN
  point-mass baseline. The dashed vertical lines corresponds to $90\%$ confidence level.}
\end{figure}

We focus now on a BNS system to study the implication of changing the PN-accuracy
of the point-mass baseline on the estimate of the tidal polarizability parameter
\begin{equation}
\tilde{\Lambda}=\dfrac{16}{13}\dfrac{(m_1+12 m_2)m_1^4 \Lambda_1 + (m_2+12 m_1)m_2^4\Lambda_2}{M^5},
\end{equation}
where $\Lambda^{i}\equiv 2/3 k^{i}_2\left[c^2/(G{{\cal C}_{i})}\right]^5$ where
${\cal C}_i\equiv m_{i}/R_{i}$ is the compactness of each star and $k_2^{i}$
the corresponding quadrupolar Love number~\cite{Damour:1983a,Hinderer:2007mb,Binnington:2009bb,Damour:2009vw}. 

We construct equal-mass EOBNR hybrid BNS waveforms by matching the {\tt TEOBResumS} EOB tidal
model~\cite{Nagar:2018zoe} to state-of-the-art NR simulations of the {\tt CoRe}
collaboration~\cite{Dietrich:2018uni}. 
Note that the version of {\tt TEOBResumS} used here does not
incorporate the analytical developments of
Refs.~\cite{Nagar:2018plt,Akcay:2018yyh}.
Two fiducial waveforms are considered here corresponding to 
two nonspinning, equal-mass ($1.35M_\odot+1.35M_\odot$) BNS models
described by the SLy and H4 EOS. The corresponding values of the tidal parameters are 
$\tilde{\Lambda}=392.231$ (Sly EOS) and $\tilde{\Lambda}=1110.5$ (H4
EOS) [For equal masses $\tilde{\Lambda}=\Lambda_1=\Lambda_2$.]
The waveforms are injected at SNR of 30 and 100 into a fiducial data stream of the
LIGO detectors~\cite{TheLIGOScientific:2014jea}.
We  assume  the  projected  noise curve  for  the  Advanced  LIGO  detectors  
in  the  zero-detuned high-power configuration  (ZDHP)~\cite{dcc:2974} 
and no actual noise is added to the data. 

\begin{table*}[t]
 \caption{Data behind Figs.~\ref{fig:sly} and \ref{fig:H4}. For each measured quantity, chirp mass ${\cal M}$, symmetric mass ratio $\nu$
 and tidal polarizability $\tilde{\Lambda}$, the colums report: the injected value; the minimum value of frequency considered, either 24Hz or 32Hz;
 the combination EOS-SNR; finally, the last three columns list the median values measured with the three different waveform approximants 
 with the $90\%$ credible interval.
 The last row of the table shows the average waveform generation time
 for each approximant, assuming starting frequency of 24Hz.}
  \label{tab:injections}
  \begin{ruledtabular}
  \begin{tabular}{cccccccc}
    & Injected Value & $f_{\rm min}$ & EOS & SNR &{\tt TaylorF2} 3.5PN &  {\tt TaylorF2} 5.5PN & {\tt IMRPhenomD\_NRTidal} \\
    \hline
    \hline
    \multirow{6}{*}{$\mathcal{M}$}  & \multirow{6}{*}{1.1752} & \multirow{4}{*}{$24$\,Hz} & SLy & 30 &   $1.1753^{1.1755}_{1.1752}$ & $1.1753^{1.1755}_{1.1752}$ & $1.1753^{1.1755}_{1.1752}$  \\
    & &  & H4 &30 &  $1.1753^{1.1755}_{1.1752}$ & $1.1753^{1.1755}_{1.1752}$ & $1.1753^{1.1755}_{1.1752}$  \\
    \cline{4-8}
    & & & SLy & 100 &  $1.1753^{1.1754}_{1.1752}$  & $1.1753^{1.1754}_{1.1752}$ & $1.1753^{1.1754}_{1.1752}$  \\
    & & &  H4 &100 & $1.1753^{1.1754}_{1.1752}$ & $1.1753^{1.1754}_{1.1752}$ & $1.1753^{1.1754}_{1.1753}$  \\
    \cline{3-8}
    & & \multirow{2}{*}{$32$\,Hz} & SLy & 30 &  $1.1754^{1.1757}_{1.1752}$ & $1.1754^{1.1757}_{1.1752}$  & $1.1754^{1.1756}_{1.1752}$ \\
    & &  & H4 &30 & $1.1754^{1.1756}_{1.1751}$  & $1.1754^{1.1757}_{1.1752}$   & $1.1753^{1.1756}_{1.1751}$  \\
    \hline
    \multirow{6}{*}{$\nu$}  & \multirow{6}{*}{0.25} & \multirow{4}{*}{$24$\,Hz} & SLy & 30 &  $0.24649^{0.24996}_{0.23147}$ &$0.24558^{0.24995}_{0.23135}$ & $0.24636^{0.24997}_{0.23105}$ \\
    & & & H4 &30 &  $0.24729^{0.24997}_{0.2338}$ & $0.24581^{0.24995}_{0.23162}$ & $0.2459^{0.24996}_{0.23079}$    \\
    \cline{4-8}
    & & & SLy & 100 & $0.24857^{0.24998}_{0.23744}$ & $0.24738^{0.24995}_{0.23691}$  & $0.24703^{0.24998}_{0.23292}$   \\
    & & & H4 &100 &  $0.24877^{0.24997}_{0.24083}$  & $0.24735^{0.24995}_{0.23694}$  & $0.24702^{0.24997}_{0.23307}$    \\
    \cline{3-8}
    & & \multirow{2}{*}{$32$\,Hz} & SLy & 30 & $0.24658^{0.24996}_{0.23107}$  & $0.2467^{0.24997}_{0.23247}$  & $0.24592^{0.24997}_{0.23054}$   \\
    & &  & H4 &30 &  $0.24684^{0.24997}_{0.23282}$  & $0.24602^{0.24995}_{0.23194}$   &  $0.24576^{0.24996}_{0.23051}$  \\
    \hline
    \multirow{6}{*}{$\tilde{\Lambda}$} & $392$ & \multirow{4}{*}{$24$\,Hz} &  SLy & 30 &  $ 935.91^{2547.71}_{245.40}$ & $ 517.88^{971.32}_{162.29}$ & $ 400.47^{761.30}_{135.47}$  \\
    & $1110$ & & H4 &30  & $ 1690.56^{3589.6}_{632.12}$ & $ 987.29^{1575.33}_{422.78}$ & $ 1044.27^{1459.18}_{630.88}$    \\
    \cline{4-8}
    & $392$ & & SLy & 100 &   $ 452.24^{694.52}_{180.44}$ & $ 301.87^{459.52}_{149.57}$ & $ 295.21^{410.31}_{162.62}$     \\
    & $1110$ &  &  H4 &100 &  $ 1405.42^{1726.90}_{1065.20}$ & $ 894.93^{1069.01}_{711.92}$ & $ 1051.61^{1195.12}_{837.18}$    \\
    \cline{3-8}
    &  $392$ & \multirow{2}{*}{$32$\,Hz} & SLy & 30 &  $ 1007.47^{2743.87}_{267.25}$ & $ 572.29^{1177.84}_{156.79}$ & $419.89^{803.14}_{144.15}$  \\
    & $1110$ &  & H4 &30 &  $ 1675.67^{3464.08}_{660.31}$ & $ 1042.61^{1713.99}_{416.23}$ & $ 1060.44^{1509.45}_{633.25}$  \\ 
    \hline
  \hline

  Average Time & & & & & 22.9 ms  &  32.68 ms & 60.13 ms \\ 
  \end{tabular}
  \end{ruledtabular}
 \end{table*}

The injected waveform is recovered with three approximants:
(i) {\tt IMRPhenomD\_NRTidal}~\cite{Dietrich:2018uni}, where the
point-mass orbital phasing is obtained by a suitable representation
of hybridized EOB/NR BBH waveforms, the {\tt PhenomD} approximant~\cite{Khan:2015jqa};
(ii) {\tt TaylorF2} where the 3.5PN orbital phase is augmented by the 6PN 
(next-to-leading) tidal phase~\cite{Vines:2010ca};
(iii) the same as above where the 3.5PN orbital, nonspinning,
phase is replaced by the quasi-5.5PN one. The models are
implemented in the ${\tt  LSC~Algorithm~Library~Suite~(LALSuite)}$. 
The LIGO-Virgo parameter-estimation algorithm {\ttfamily LALInference}~\cite{Veitch:2014wba}) 
is then employed to extract the binary properties from the signal.
We use a uniform prior distribution in the interval $[1 M_\odot, 3 M_\odot]$ for the
component masses, and a uniform prior between $-1$ and $1$ 
for both dimensionless aligned spins.  We also pick a uniform 
prior distribution for the individual tidal parameters
$\Lambda_{1,2}$ between $0$ and $5000$.

The outcome of the analysis is illustrated in Fig.~\ref{fig:sly} for the SLy
EOS and  and Fig.~\ref{fig:H4} the H4 EOS. We compare 
the inference of the tidal parameter done on two frequency 
intervals, $[24,1024]$~Hz and $[32,1024]$. Note that we
do not extend the analysis interval even further because we
know that the orbital part of the {\tt TaylorF2} approximants
becomes  largely inaccurate at higher frequencies.
For ${\rm SNR}=30$ one finds that the 
3.5PN orbital baseline induces a clear bias in $\tilde{\Lambda}$, while the 
quasi-5.5PN one agrees much better with the {\tt PhenomD} model as well
as the expected value (vertical line in the plots). 
Incrementing the SNR to 100, the statement only holds for 
the softer  EOS, since for the H4 case also the 5.5PN 
approximant is biased,  although still less than the 3.5PN one.
The two figures are complemented by Table~\ref{tab:injections},
that, for each choice of configuration and SNR, lists the recovered
values with their 90\% credible interval.
The last row of the table also reports the time needed to generate
a single waveform during the PE process: interestingly, the timing of the 
quasi-5.5PN {\tt TaylorF2} is comparable to the one of the 3.5PN approximant,
i.e. it remains approximately {\it two times faster} than {\tt PhenomD\_NRTidal}
being consistent with this latter at SNR $\lesssim30$. 
This suggests that, for events similar to GW170817 or quieter, the quasi-5.5PN 
{\tt TaylorF2} can effectively be used in place of {\tt PhenomD\_NRTidal} 
to get an even faster, yet accurate, estimate of the parameters.

\subsection{Understanding waveform systematics of the injections via the $Q_\omega$ analysis}

%
%
\begin{figure}[t]
\begin{center}
\includegraphics[width=0.9\columnwidth]{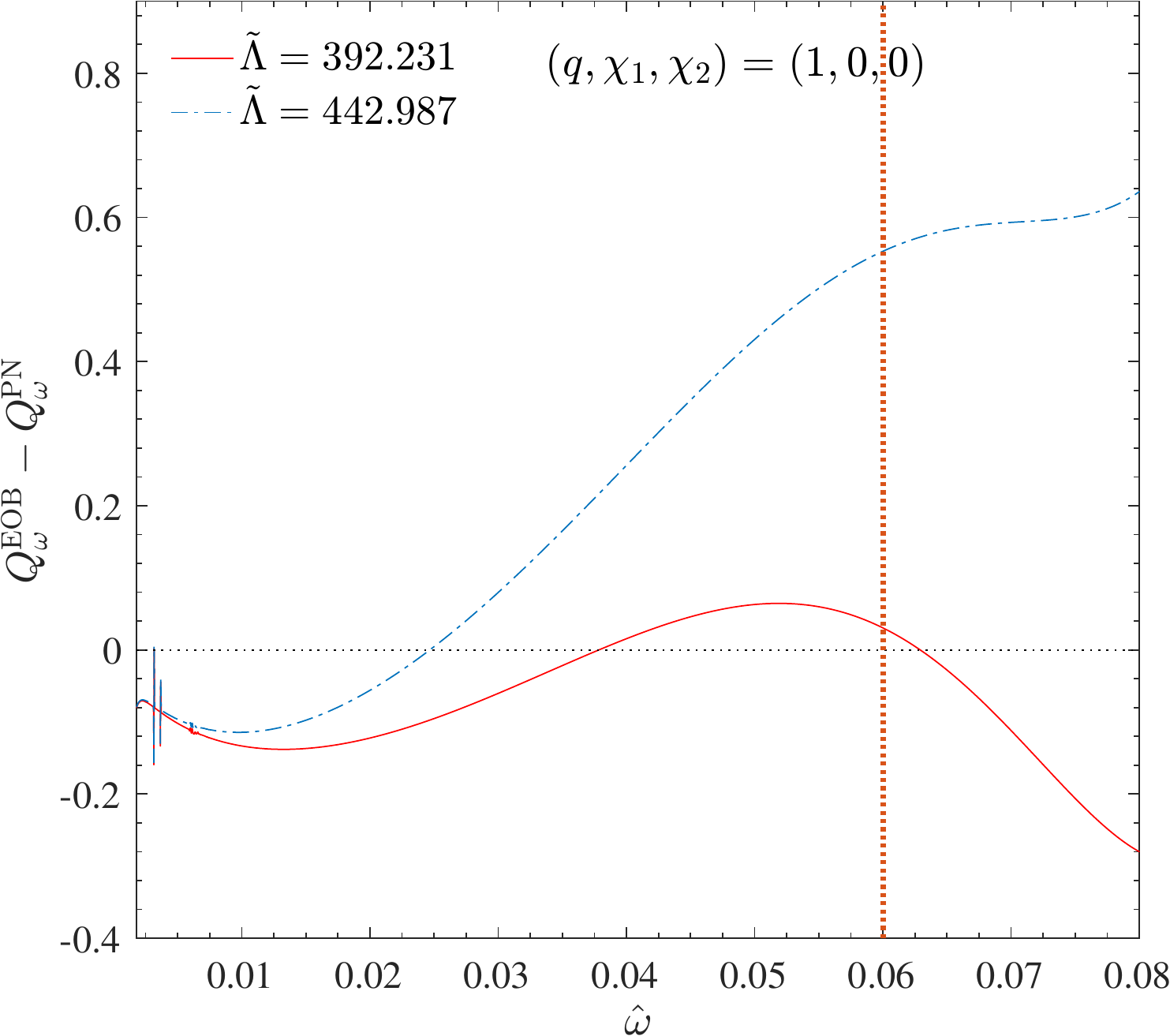}
\end{center}
\caption{ \label{fig:bias_explanation} Heuristic explanation of
  the bias on $\tilde{\Lambda}$: $1.35M_\odot+1.35M_\odot$ binary,
  Sly EOS, $\tilde{\Lambda}=392.231$. Shown is the gauge-invariant difference
  $\Delta Q_{\omega}^{\rm EOBPN}\equiv Q_\omega^{\rm EOB}-Q_\omega^{\rm PN}$ between
  the EOB $Q_\omega$ and the PN $Q_\omega$ with the 3.5PN orbital baseline augmented
  by the 6PN-accurate tidal phase. Increasing the value of the tidal parameter to $\tilde{\Lambda}=442.987$
  is very effective in reducing the phase difference accumulated between in the interval
  $\omega\in [0.02,0.06]$ (dotted vertical line) to a negligible value.
  Such $\hat{\omega}$ interval correponds to $f\in [24,718]$~Hz for this binary.
  The upper frequency limit corresponds to 957.4~Hz.}
\end{figure}

Let us finally heuristically explain why the effect of the
3.5PN-accurate orbital baseline is to bias the value of $\tilde{\Lambda}$
towards values that are larger than the theoretical expectation.
Inspecting Fig.~\ref{fig:Qwcomp_q1} one sees that the
$Q_\omega^{\rm EOB}-Q_\omega^{\rm 3.5PN}$ is negative.
This means that the PN phase accelerates {\it less} than the EOB one,
namely the inspiral occurs {\it more slowly} in the 3.5PN phasing
description than in the EOB one. Loosely speaking, one may think
that the gravitational interaction behind the 3.5PN-accurate orbital phasing is
{\it less attractive} than what predicted by the EOB model.
Evidently, this effect might be compensated by and additional part
in the total PN phasing that stems for a part of the dynamics that
is intrinsically {\it attractive} and that could compensate for the
inaccurate behavior of the 3.5~PN. Since eventually the phase difference
is given by an integral, two effects of opposite sign can mutually
compensate and thus generate a PN-based frequency phase that is
compatible with the EOB one. Since tidal interactions are attractive,
the corresponding part of the phasing is naturally able to compensate
the repulsive character of the orbital phasing.
For this compensation to be effective, it may happen that
$\tilde{\Lambda}$ has to be {\it larger} than the theoretically
correct one that accounts for the tidal interaction (at leading order)
in the EOB waveform. 

Such intuitive explanation is put on more solid
ground in Fig.~\ref{fig:bias_explanation}.
The figure refers to the SLy model and compares two EOB-PN $Q_\omega$
differences $\Delta Q_{\omega}^{\rm EOBPN}\equiv Q_\omega^{\rm EOB}-Q_\omega^{\rm PN}$,
where the $Q_\omega^{\rm EOB}$ is the complete function, while $Q_\omega^{\rm PN}$
is obtained summing together the 3.5PN orbital phase and the 6PN-accurate
tidal phase~\cite{Vines:2011ud}. When we use the theoretically correct
value of 
$\tilde{\Lambda}=\Lambda_1=\Lambda_2=392.231$ the phase
difference
in the interval $\hat{\omega}\in [\hat{\omega}_0,\hat{\omega}_1]=[0.002,0.06]$,
corresponding to $f\in[20,718]$~Hz (dotted vertical line in the figure)
for this binary, is $\Delta\phi_{(\hat{\omega}_0,\hat{\omega}_1)}\simeq -0.276$~rad. By contrast, if the value
of $\tilde{\Lambda}$ is progressively increased, the accumulated phase
difference between  $[\hat{\omega}_0,\hat{\omega}_1]$ gets reduced up
to $\Delta\phi_{(\hat{\omega}_0,\hat{\omega}_1)}\simeq 2.429\times 10^{-4}$ for $\tilde{\Lambda}=442.987$. Note however
that such analytically predicted ``bias''in $\tilde{\Lambda}$ depends
on the frequency interval considered: if we extended the integration
up to $\hat{\omega}_1\simeq 0.08$ (corresponding to 957.4~Hz) one finds
that a similarly small accumulated phase difference $\Delta\phi_{0.002,0.08}\simeq 5.0\times 10^{-5}$
is obtained for $\tilde{\Lambda}=424.08$, i.e. the analytical bias is {\it reduced}.
This fact looks counterintuitive: a result obtained with a PN approximant
is not, a priori, expected to improve when including higher frequencies.
By contrast, the fact that the analytic bias is (slightly) reduced increasing
$\hat{\omega}_1$ just illustrates the {\it lack of robustness} as well
as the {\it lack of predictive power} of the approximant in the strong-field regime.
Generally speaking, one sees that the combination of 3.5PN orbital phase
with 6PN tidal phase may result in a waveform that is {\it effectual} with
respect the EOB one, in the sense that the noise-weighted scalar product
will be of order unity, but with an incorrect value of the tidal parameter.
This simple example is helpful to intuitively understand how the incorrect
behavior of the point-mass nonspinning phasing can eventually result
in a bias towards larger values of $\tilde{\Lambda}$.
Interestingly, this value is close to the value obtained with SNR=100
(see left-bottom panel of Fig.~\ref{fig:sly}). Although the analysis of
Fig.~\ref{fig:bias_explanation} certainly cannot replace an injection-recovery
study, it should be kept in mind as a complementary tool to interpret its
outcome within a simple, intuitive but quantitative, framework.

\section{Conclusions}
\label{sec:conclusions}

Our results can be summarized as follows:
\begin{enumerate}
\item Starting from the EOB resummed expressions for the energy flux and energy along circular orbits, we have computed a
  {\tt TaylorF2} point-mass, nonspinning, approximant at formal 5.5PN order. Such quasi-5.5PN approximant depends
  on some, yet uncalculated, $\nu$-dependent, PN parameters that are
  set to zero.
  
\item Among various truncations of the 5.5PN approximant (3.5PN, 4PN,
  4.5PN, 5PN, see Appendix~\ref{sec:why_5p5PN}, we have found that the 5.5PN phasing
  performs best when compared with the complete point-mass phasing obtained with ${\tt TEOBResumS}$. Such phasing
  comparison was done exploiting its gauge-invariant description through the $Q_\omega=\hat{\omega}^2/\dot{\hat{\omega}}$
  function. The main outcome of this analysis is that the EOB-derived quasi-5.5PN approximant
  is {\it remakably close} to the complete EOB phasing up to the late inspiral (e.g. $\hat{\omega}\simeq 0.06$)
  and performs better than the standard (analytically complete) 3.5PN
  one. We tested that the performaces remain robust for unequal masses up to $q=2$
  and aligned spin cases with dimensionless spin magnitudes up to $\chi\sim\pm0.1$.

\item To assess the use of the 5.5PN approximant in GW parameter
  estimations, we considered injection studies with hybrid waveforms of GW170817-like sources.
  The improved {\tt TaylorF2} point-mass baseline reduces (or even
  eliminates) the biases on the measurability of the tidal
  polarizability parameter $\tilde{\Lambda}$ instead 
  produced by the use the standard 3.5PN point-mass baseline.
  Therefore, the new 5.5PN approximant can be faithfully and
  effectively used in matched filtered
  searches and Bayesian parameter estimation. 
 
\end{enumerate}

We recommend to use the new quasi-5.5PN approximant to improve the
performance of the  {\tt TaylorF2} and substitute it to the 3.5PN in
searches/parameter estimation. To ease this task, we have implemented
all the new PN terms up to 5.5PN in the ${\tt LSC~Algorithm~Library~Suite~(LALSuite)}$.

The performances of the 5.5PN approximant could be further improved 
towards higher frequencies by carefully tuning some of the free parameters. 
A preliminary investigation based on a equal masses nonspinning BNS is 
presented at the end of Appendix~\ref{sec:why_5p5PN}). We find that by tuning 
the 5PN parameter $a_6^c$ and the 4PN coefficient $c^{\rm 4PN}_{22}$ entering the $\ell=m=2$ 
waveform amplitude, the phase difference between such flexed PN approximant and
the EOB phasing is negligible essentially up to merger. This indicates that, while 
the PN series keeps oscillating even when high order terms come into play, 
future work might be devoted to effectively minimize such oscillations by suitably
tuning such parameters.

\acknowledgments
We are grateful to T.~Damour for discussions and comments at the very beginning of this work,
and for a careful reading of the manuscript. The original ideas that eventually led to
this study were elaborated in discussions between T.~Damour, A.~N. and L.~Villain.
In this respect, A.~N.  especially acknowledges discussions with L.~Villain that
helped identifying several technical problems related to the efficient generation 
of EOB waveforms from  the early-frequency regime. We also thank S.~Khan and A.~Samajdar for helping in 
implementing the approximant in ${\tt LALSuite}$. F.~M.~thanks IHES for hospitality
at various stages during the development of this work. R.~D. was supported in part by DFG
grants GK 1523/2 and BR 2176/5-1. S.~B. acknowledges support by
the EU H2020 under ERC Starting Grant, no.~BinGraSp-714626. 

\appendix

\begin{figure}[t]
\center
  \includegraphics[width=0.99\columnwidth]{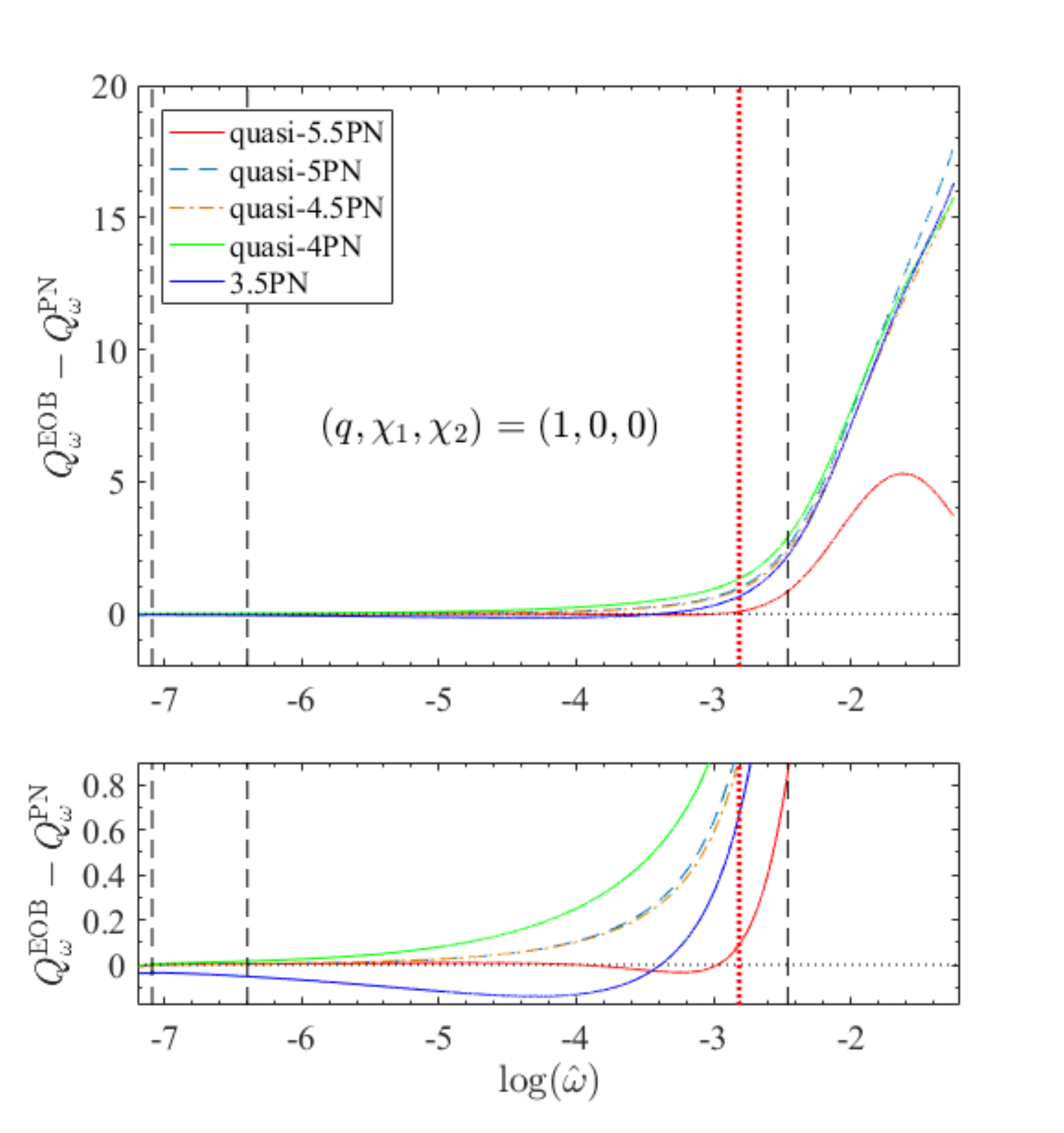} 
  \caption{ \label{fig:Qwcomp_q1_ALL} Comparison between the point-mass (nonspinning) orbital phasing for $q=1$
$Q_{\omega}^{\rm EOB}-Q_{\omega}^{\rm PN}$ difference up to (approximate) merger time. The vertical lines mark
the 10Hz, 20Hz, 718Hz or 1024Hz for a  $(1.35+1.35)M_{\odot}$ binary. The quasi-5.5PN curve is always
much closer to the EOB one than the other PN order approximants. See Table \ref{tab:DphiALL} for the accumulated phase differences.}
\end{figure}

\begin{table}
  \caption{\label{tab:DphiALL} EOB/PN phase difference accumulated between $[f_0,f_1]$. It is 
  obtained by integrating the $\Delta Q_\omega^{\rm EOBPN}$'s in Fig.~\ref{fig:Qwcomp_q1_ALL} 
  between the corresponding values of $\log(\hat{\omega})$, also listed in the table. 
  The limits of integration denoted in Hz refer to the fiducial $(1.35+1.35)M_\odot$ binary system.}
  \begin{ruledtabular}
    \begin{tabular}{cccccccc}
      $\hat\omega_0$ $\times 10^{4}$& $\hat\omega_1$ & $f_0$[Hz]  & $f_1$[Hz]  & $\Delta\phi^{\rm EOBPN}_{\rm 5PN}$ & $\Delta\phi^{\rm EOBPN}_{\rm 4.5PN}$ & $\Delta\phi^{\rm EOBPN}_{\rm 4PN}$\\
      \hline
      \hline
     $8.35$ &0.086&10 & 1024 & 1.0805 & 1.0109 & 1.6306\\
     $8.35$ & 0.060 &10 & 718   & 0.4984 & 0.469 & 0.9241\\
      \hline
      $16.7$ & 0.086 &20 & 1024 & 1.079 & 1.0094 & 1.6235\\
      $16.7$ & 0.060 &20 & 718   & 0.4970 & 0.4675 & 0.9170\\
      \hline
      $20.0$ & 0.086 &24 & 1024 & 1.0781  & 1.0085 & 1.6203 \\            
    \end{tabular}
  \end{ruledtabular}
  \end{table}
  
\begin{figure}[t]
\center
  \includegraphics[width=0.90\columnwidth]{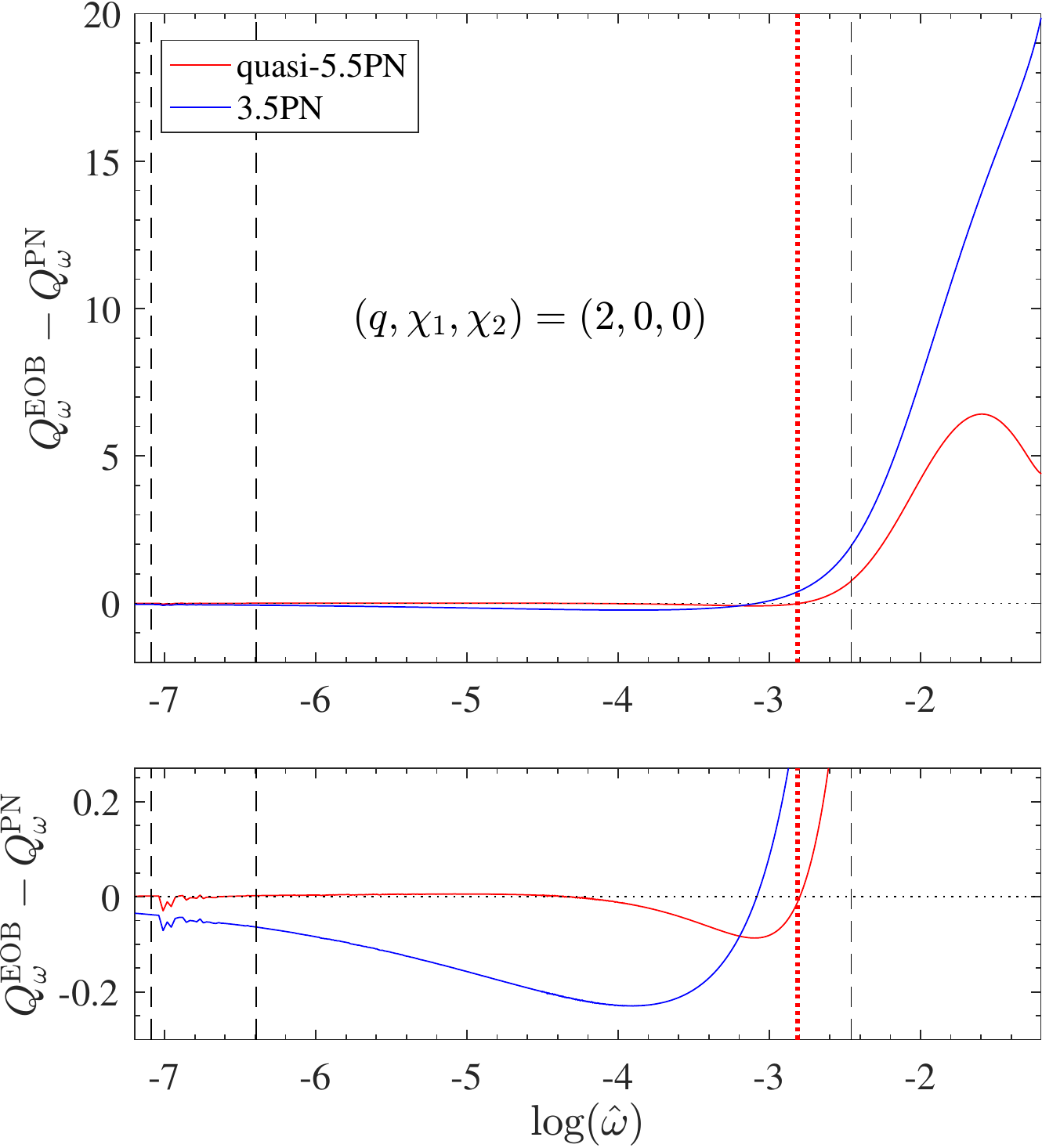}
  \caption{ \label{fig:Qwcomp_q2} Robustness of the quasi-5.5PN {\tt TaylorF2} approximant versus mass ratio. 
  The vertical lines correspond to the same four values of $[\hat{\omega}_1,\hat{\omega}_2]$ listed in
  the first two columns of Table~\ref{tab:DphiALL}.}
\end{figure}

\begin{figure}[t]
\center
  \includegraphics[width=0.75\columnwidth]{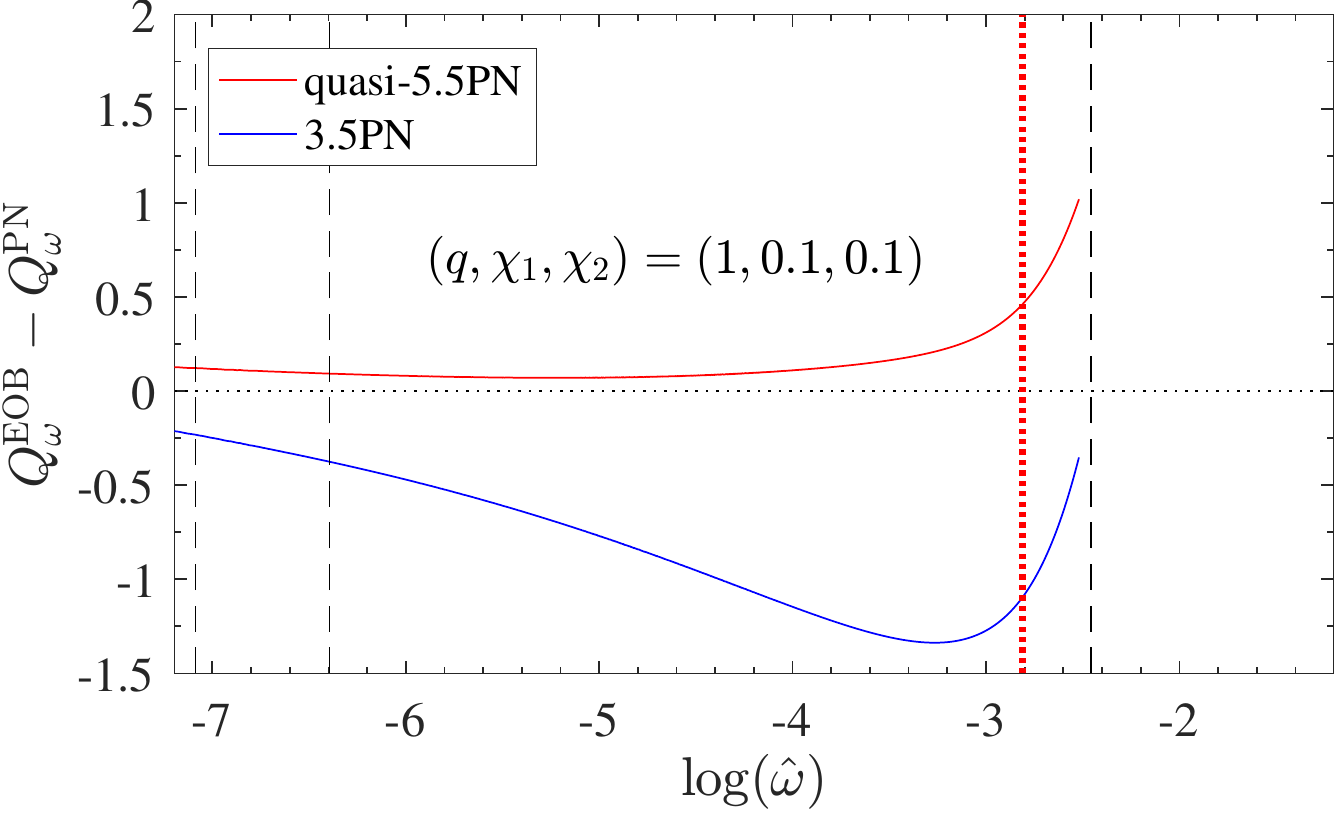}\\
  \includegraphics[width=0.75\columnwidth]{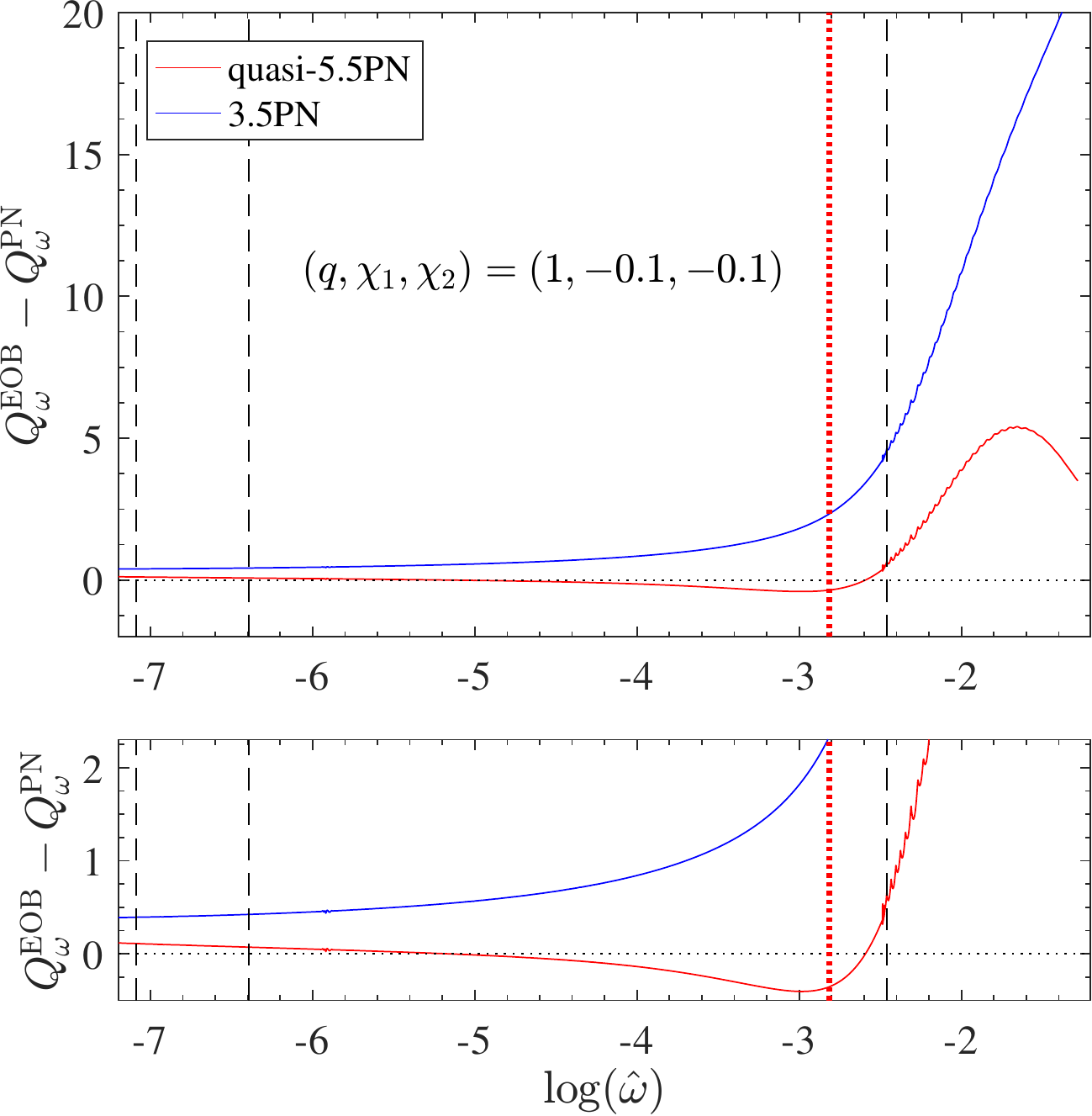}
  \caption{ \label{fig:Qwcomp_q1_s0101}Robustness of the quasi-5.5PN {\tt TaylorF2} approximant versus spin:
    spins aligned (top) and spins anti-aligned (bottom) with the orbital angular momentum. The EOB/PN agreement 
    is always improved by the use of the quasi-5.5PN orbital phasing.  The vertical lines correspond to the same 
    four values of $[\hat{\omega}_1,\hat{\omega}_2]$ listed in the first two columns of Table~\ref{tab:DphiALL}.}
\end{figure}

\section{Why quasi-5.5PN?}
\label{sec:why_5p5PN}
Post-Newtonian expansion are truncated asymptotic series, so it  is not a priori granted that by increasing the 
PN order one will automatically get a better approximation to the exact result. The choice of using the quasi-5.5PN 
${\tt TaylorF2}$ for the injection study of Sec.~\ref{sec:DA} was made after having carefully analyzed 
all the previous quasi-PN orders beyond 3.5PN and having compared each PN-truncation of the 
$Q_\omega$ function to the corresponding outcome of  {\tt TEOBResumS}.
The result of this analysis is shown in Fig.~\ref{fig:Qwcomp_q1_ALL}, that illustrates how the quasi-5.5PN
$Q_{\omega}^{\rm EOB}-Q_{\omega}^{\rm PN}$ difference remains consistently close to zero for a frequency
interval that is much longer than for any other lower-PN truncation. This finding justifies our choice of
focalizing specifically on the quasi-5.5PN approximant in the main text.

With this PN order, the first natural question that follows is whether there is some simple
way to improve the accuracy of the approximant just by tuning some of its (many) free parameters.
Before entering this discussion, the simplest thing to do is to incorporate more analytical
information ,e.g. instead of using $a_6^c=0$ incorporating either the analytical gravitational
self-force value $a_6^c(0)$, that was obtained in Refs.~\cite{Barausse:2011dq,Bini:2013rfa}
\begin{align}
a_{6}^c(0)&=-\frac{1066621}{1575}-\frac{14008 \gamma }{105}+\frac{246367 \pi ^2}{3072}\nonumber\\
&-\frac{31736 \log (2)}{105}+\frac{243 \log(3)}{7},
\end{align}
or even the numerical-relativity informed one~\cite{Nagar:2015xqa}
\begin{equation}
a_{6,\rm NR}^c(\nu)=3097.3 \nu ^2-1330.6 \nu +81.38.
\end{equation}
Although these two functions encode physically correct effect (though only effectively for $a_{6,\rm NR}^c(\nu)$)
they turn out, both, to increase the repulsive character of the approximant, without any real advantage.
In practice, when the values above are used, one get an acceptable EOB/PN agreement only
up to $\hat{\omega} \approx 0.02$.

From the PN point of view, we expect from Table II of~\cite{Messina:2018ghh} that the order
of magnitude of the various $c_{\ell m}^{\rm PN}$ coefficient is small. This is the rational
behind our conservative choice of simply setting then to zero. Still, one could think to use
some of these coefficients, as well as $a_6^c$ as tunable parameter and investigate whether
it is possible to flex the quasi-5.5PN {\tt TaylorF2} approximant so as to reduce the
EOB/PN disagreement even to frequencies higher than $0.06$.

As a proof of principle, we explored that this is the case considering the $q=1$
case and flexing, at the same time, both $a_6^c$ and $c_{22}^{\rm 4PN}$, that are, in sense,
the lowest-order unknown coefficients in the model.
One easily finds that fixing $a_6^c=49$ and $c_{22}^{\rm 4PN}=10.45$ the integrated
EOB/PN phase difference in the point-mass sector accumulated on the interval
$[\hat\omega_0,\hat\omega_1]=[0.002,0.08]$ can be reduced from $0.085$~rad to $\simeq 10^{-4}$ rad.

\section{Mass ratio and spin}
\label{sec:mass_ratio}
Increasing the mass ratio or the spins (to mild values) of the binary does not affect the robustness of the quasi-5.5PN, 
untuned, approximant. In Fig.~\ref{fig:Qwcomp_q2} we see that the difference between the ``exact" 
numerical $Q_{\omega}$ and the analytical one is still approximately flat for a qualitatively wide range 
of frequencies. Beyond that, for what concerns the spin, we show in Figs.~\ref{fig:Qwcomp_q1_s0101} a qualitative point-mass BNS case with realistic positive and 
negative spins. In this case we put to zero the quadrupole-monopole interaction terms, $C_{Q1}=C_{Q2}=0$, 
removing the quadratic-in-spin PN corrections both in the numerical ${\tt TEOBResumS}$ model and in the 
analytical approximant. So contribution of the mixed $\chi_1\chi_2$ terms and the spin-orbit interaction 
is tested, including for completeness also the new 4PN spin-orbit {\tt Taylor-F2} term computed in 
Ref.~\cite{Nagar:2018plt}, i.e. the $Q_{\omega}$ analogue of Eq.~(48) there. While the spin-orbit terms 
are already contained in ${\tt TEOBResumS}$ in a resummed form, we neglect the spin cube and spin 
quartic PN corrections (see \cite{Nagar:2018plt}) for simplicity, since their effect does not affect our 
preliminary robustness test.

\section{Quasi-5.5PN phasing coefficients}
\label{sec:coefficients}
We report in this appendix the explicit expressions for the coefficients entering Eq.~\eqref{eq:Qw5p5pn}.
For simplicity, we put to zero all the $c_{\ell m}^{\rm PN}$'s except $c_{22}^{\rm 4PN}$ and  $c_{21}^{\rm 3PN}$.
We have:
\begin{widetext}
\begin{align}
\nonumber
b_2&=\frac{11 \nu }{4}+\frac{743}{336}\\
b_3&=-4 \pi\\
b_4&=\frac{617 \nu ^2}{144}+\frac{5429 \nu }{1008}+\frac{3058673}{1016064}\\
b_5&=\pi\left(\frac{13 \nu }{8}-\frac{7729}{672}\right)\\
b_6&=\frac{25565 \nu ^3}{5184}-\frac{15211 \nu ^2}{6912}+\left(-\frac{451 \pi^2  }{48}
    +\frac{3147553127}{12192768}\right)\nu\nonumber\\
   &+\frac{856 \log (x)}{105}+\frac{3424 \log (2)}{105}+\frac{32 \pi
   ^2}{3}+\frac{1712 \gamma }{105}-\frac{10817850546611}{93884313600}\\
b_7&=\pi\left(\frac{14809  \nu ^2}{3024}-\frac{75703  \nu
}{6048}-\frac{15419335 }{1016064}\right)\\
b_8&=c_{21}^{\rm 3PN}\biggl(\frac{4}{9}\nu^2-\frac{1}{9}\nu\biggr)+\frac{73893895655 \nu ^4}{14239120896}-\frac{102008296205\nu ^3}{11650189824}
   +\biggl(\frac{79909 \pi ^2}{24192}-\frac{300600673165997}{2563041761280}\biggr)\nu^2\nonumber\\
   &+\biggl(-4 c_{22}^{\rm 4PN}  +\frac{332683 \log (x) }{2205}-\frac{1860443 \pi ^2 }{48384}
   +\frac{6252765829282087}{5695648358400}\biggr)\nu\nonumber\\
   &+\gamma  \biggl(\frac{665366 \nu
   }{2205}+\frac{9203}{210}\biggr)+\frac{9203 \log (x)}{420}+\biggl(\frac{47385}{1568}-\frac{47385 \nu }{392}\biggr)
   \log(3)\nonumber\\
   &+\biggl(\frac{177586 \nu }{245}+\frac{50551}{882}\biggr) \log (2)+\frac{9049 \pi
     ^2}{252}-\frac{2496799162103891233}{3690780136243200}\,
\end{align}
\begin{align}
b_9&= \pi\bigg[\frac{2064751  \nu^3}{399168}+\frac{9058667   \nu^2}{254016}+\biggl(\frac{451 \pi ^2}{12}-\frac{298583452147
   }{268240896}\biggr)\nu\nonumber\\
   &-\frac{3424}{105}   \log (x)-\frac{13696}{105}   \log (2)-\frac{64 \pi ^2}{3}-\frac{6848 \gamma}{105}+\frac{90036665674763  }{187768627200}\bigg]\,,\\
b_{10}&=\frac{1}{-1+3\nu}\bigg[c_{21}^{\rm 3PN}\biggl(\frac{121}{189}\nu-\frac{2815 }{756}\nu^2+\frac{599}{252}\nu^3+\frac{191}{21}\nu^4\biggr)+ \biggl(-12 c_{22}^{\rm 5PN}+54 a_6^c+\frac{349 c_{22}^{\rm 4PN}}{42}\nonumber\\
&-\frac{415795517 \pi
   ^2}{3612672}+\frac{99239192119 \gamma }{18336780}+\frac{3592581310185992768897549}{292602096577251901440}+\frac{58330935 \log
   (3)}{21952}\nonumber\\
   &+\frac{117340379713 \log (2)}{18336780}\biggr)\nu^2+  \biggl(-18 a_6^c +\frac{311 c_{22}^{\rm 4PN}}{42}+4 c_{22}^{\rm 5PN}+\frac{5451429547 \pi ^2}{32514048}\nonumber\\
   &-\frac{63787407527 \gamma
   }{36673560}-\frac{2749231177355819921781277}{216742293760927334400}-\frac{8924175 \log (3)}{12544}
   -\frac{159943397077 \log
   (2)}{73347120}\biggr)\nu\nonumber\\
   &+ \biggl(-\frac{641 c_{22}^{\rm 4PN} }{7}+\frac{460519 \pi ^2}{6144}+\frac{3073896571 \gamma
   }{3056130}+\frac{655042586669421296014259}{137157232770586828800}\nonumber\\
   &-\frac{1279395 \log (3)}{392}+\frac{10777949417 \log
   (2)}{1528065}\biggr)\nu^3-\frac{2425066585102052979797 \nu ^6}{3428930819264670720}+\frac{12721434740371951621 \nu
   ^5}{2705270863325184}\nonumber\\
   &+\left(\frac{9434797 \pi ^2}{75264}-\frac{11160810800663155149913}{1088549466433228800}\right) \nu
   ^4-\frac{578223115 \pi ^2}{12192768}-\frac{6470582647 \gamma
   }{110020680}\nonumber\\
   &+\frac{1412206995432957982751}{505226791983513600}+\frac{5512455 \log (3)}{87808}-\frac{53992839431 \log(2)}{220041360}\bigg]\nonumber\\
  &+\left(-\frac{6431890181 \nu^2 }{18336780}-\frac{2968141499\nu}{12224520}+\frac{6470582647}{220041360}\right) \log (x)\,,\\
b_{11}&=\pi\bigg[c_{21}^{\rm 3PN}\biggl(-\frac{8}{3}\nu^2+\frac{2}{3}\nu\biggr)+ \frac{131525414689  \nu ^4}{64076044032}-\frac{216119565695   \nu ^3}{11650189824}+\biggl(-\frac{2129581 \pi ^2 
  }{16128}  +\frac{5120314955146397 }{1398022778880}\biggr)\nu ^2\nonumber\\
   &+\biggl(-\frac{673331   \log (x) }{1260}
   +\frac{9003157 \pi ^2  }{64512}+16 c_{22}^{\rm 4PN} 
   -\frac{471473599592788087  }{76891252838400}\biggr)\nu\nonumber\\
   &+\gamma  \biggl(-\frac{673331   \nu }{630}-\frac{3558011 
   }{17640}\biggr)-\frac{3558011   \log (x)}{35280}+\biggl(\frac{47385   \nu }{196}-\frac{47385  }{784}\biggr) \log
   (3)\nonumber\\
   &+\biggl(-\frac{10504813   \nu }{4410}-\frac{862549  }{2520}\biggr) \log (2)-\frac{9439
   \pi^2}{126}+\frac{1795505143426433771  }{615130022707200}\bigg] \ .
\end{align}
\end{widetext}
The GW phase in the SPA is computed from the $Q_\omega$ using
\eqref{eq:Psi} and it is given by the Taylor series
\begin{align}
 \label{eq:PsiTaylor}
 \Psi(f) & = \frac{3 (\pi M f)^{-5/3}}{128\nu}\sum_i \varphi_i (\pi M
 f)^i \ ;
\end{align}
with the coefficients:
\begin{widetext}
\begin{align}
\varphi_0&=1\\\nonumber
\varphi_1&=0\\\nonumber
\varphi_2&=\frac{3715}{756}+\frac{55}{9}\nu\\\nonumber
\varphi_3&=-16\pi\\\nonumber
\varphi_4&=\frac{3085 \nu ^2}{72}+\frac{27145 \nu }{504}+\frac{15293365}{508032}\\\nonumber
\varphi_5&=\pi\left(\frac{38645}{756}-\frac{65 \nu }{9}\right) [1+\log (\pi  f M)]\\\nonumber
\varphi_6&=-\frac{6848}{63} \log (\pi  f M)-\frac{127825 \nu ^3}{1296}+\frac{76055 \nu ^2}{1728}+\biggl(\frac{2255 \pi ^2
   }{12}-\frac{15737765635}{3048192}\biggr)\nu-\frac{640 \pi ^2}{3}-\frac{6848 \gamma}{21}\\\nonumber
   &+\frac{11583231236531}{4694215680}-\frac{13696 \log (2)}{21}\\\nonumber
\varphi_7&=\pi\biggl(-\frac{74045 \nu ^2}{756}+\frac{378515 \nu }{1512}+\frac{77096675}{254016}\biggr)\\\nonumber
\varphi_8&= [1-\log (\pi f M)] \biggl[c_{21}^{\rm 3PN} \biggl(\frac{40 \nu }{81}-\frac{160 \nu ^2}{81}\biggr)+\frac{160 c_{22}^{\rm 4PN} \nu
   }{9}-\frac{369469478275 \nu ^4}{16019011008}+\frac{510041481025 \nu
   ^3}{13106463552}\\\nonumber
   &+\biggl(\frac{300600673165997}{576684396288}-\frac{399545 \pi ^2}{27216}\biggr) \nu ^2+
   \biggl(-\frac{5679872289503527}{1281520880640}-\frac{5322928 \gamma }{3969}+\frac{9302215 \pi ^2}{54432}\\\nonumber
   &-\frac{1420688 \log
   (2)}{441}+\frac{26325 \log (3)}{49}\biggr)\nu-\frac{90490 \pi ^2}{567}-\frac{36812 \gamma
   }{189}+\frac{2550713843998885153}{830425530654720}-\frac{26325 \log (3)}{196}\\\nonumber
   &-\frac{1011020 \log
   (2)}{3969}\biggr]+\biggl(\frac{2661464 \nu }{11907}+\frac{18406}{567}\biggr)\log^2 (\pi f M)  \\\nonumber
\varphi_9&=\pi  \biggl[-\frac{13696}{63} \log (\pi  f M)+\frac{10323755 \nu ^3}{199584}+\frac{45293335 \nu ^2}{127008}
   +\biggl(\frac{2255  }{6}\pi^2-\frac{1492917260735 
   }{134120448}\biggr)\nu\\\nonumber
   &-\frac{640}{3}\pi^2-\frac{13696 \gamma
   }{21}+\frac{105344279473163}{18776862720}-\frac{27392 \log (2)}{21}\biggr]
\end{align}
\end{widetext}
\begin{widetext}
\begin{align}
\varphi_{10}&=\frac{1}{1-3\nu}\biggl[a_6^c(72\nu-216\nu^2)+c_{21}^{\rm 3PN} \left(-\frac{764 \nu ^4}{21}-\frac{599 \nu ^3}{63}+\frac{2815 \nu ^2}{189}-\frac{484 \nu
   }{189}\right)
   +c_{22}^{\rm 4PN} \left(\frac{2564 \nu ^3}{7}-\frac{698 \nu ^2}{21}-\frac{622 \nu }{21}\right)\\\nonumber
   &+c_{22}^{\rm 5PN}
   \left(48 \nu ^2-16 \nu \right)+\left(\frac{12863780362 \nu ^3}{4584195}+\frac{13849493129 \nu ^2}{13752585}-\frac{24279431641
   \nu }{27505170}+\frac{6470582647}{82515510}\right) \log (\pi  f M)\\\nonumber
   &+\frac{2425066585102052979797 \nu
   ^6}{857232704816167680}-\frac{12721434740371951621 \nu
   ^5}{676317715831296}+\biggl(\frac{11160810800663155149913}{272137366608307200}\\\nonumber
   &-\frac{9434797 \pi ^2}{18816}\biggr) \nu ^4+ \biggl(-\frac{857104076559310860540851}{34289308192646707200}-\frac{6147793142 \gamma }{1528065}-\frac{460519 \pi
   ^2}{1536}\\\nonumber
   &-\frac{43111797668 \log (2)}{1528065}+\frac{1279395 \log (3)}{98}\biggr)\nu^3+
   \biggl(-\frac{18736399363805057301105217}{365752620721564876800}-\frac{99239192119 \gamma }{4584195}\\\nonumber
   &+\frac{415795517 \pi
   ^2}{903168}-\frac{117340379713 \log (2)}{4584195}-\frac{58330935 \log (3)}{5488}\biggr)\nu^2+ 
   \biggl(\frac{569935181259668744781113}{10837114688046366720}\\\nonumber
   &+\frac{63787407527 \gamma }{9168390}-\frac{5451429547 \pi
   ^2}{8128512}+\frac{159943397077 \log (2)}{18336780}+\frac{8924175 \log (3)}{3136}\biggr)\nu+\frac{578223115 \pi
   ^2}{3048192}\\\nonumber
   &+\frac{6470582647 \gamma }{27505170}-\frac{1433006523295407126559}{126306697995878400}-\frac{5512455 \log
     (3)}{21952}+\frac{53992839431 \log (2)}{55010340}\biggr]\\\nonumber
\varphi_{11}&=\pi  \biggl[ c_{21}^{\rm 3PN}\biggl(-\frac{160 \nu ^2}{27}+\frac{40  \nu }{27}\biggr)+\frac{320 c_{22}^{\rm 4PN} \nu
   }{9}-\biggl(\frac{1346662 \nu }{1701}+\frac{3558011}{23814}\biggr) \log (\pi  f M)\\\nonumber
   &+\frac{657627073445 \nu
   ^4}{144171099072}-\frac{1080597828475 \nu ^3}{26212927104}+\frac{5120314955146397 \nu ^2}{629110250496}+\pi ^2
   \biggl(-\frac{10647905 \nu ^2}{36288}+\frac{45015785 \nu }{145152}\\\nonumber
   &-\frac{94390}{567}\biggr)+\biggl(-\frac{1346662 \gamma  
   }{567}-\frac{430383707398397047  }{34601063777280}+\frac{26325}{49}   \log (3)-\frac{21009626   \log
   (2)}{3969}\biggr)\nu-\frac{3558011 \gamma }{7938}\\\nonumber
   &+\frac{1857541407236594411}{276808510218240}-\frac{26325 \log (3)}{196}-\frac{862549
   \log (2)}{1134}\biggr]   
\end{align}
\end{widetext}

\bibliography{refs20190421.bib}

\end{document}